\documentclass{aa}
\usepackage{graphicx}

\begin{document}
\title{Chemical evolution and abundance gradients in the Milky Way}

\author{Andreu Alib\'es\inst{1}
  \and Javier Labay\inst{1}
  \and Ramon Canal\inst{1,2}}

\offprints{A. Alib\'es, aalibes@am.ub.es}

\institute{Departament d'Astronomia i Meteorologia, Universitat de Barcelona,
           Mart\'{\i} i Franqu\`es 1, 08028 Barcelona, Spain
	   \and Institut d'Estudis Espacials de Catalunya, Edifici Nexus, Gran
	   Capit\`a 2-4, 08034 Barcelona, Spain}

\date{Received / Accepted}
\authorrunning{Alib\'es, Labay, \& Canal}
\titlerunning{Abundance gradients in the Milky Way}

\abstract{We extend our model of chemical evolution that successfully account for the main observables in the solar neighborhood (Alib\'es, Labay \& Canal \cite{alib01}) to the whole Milky Way halo and disk. We assume an inside--out scenario for the assembling of the Galaxy through the infall of external material. Two different compositions of the infalling material have been considered: primordial and slightly metal--enriched ($0.1\, Z_{\sun}$). Our calculations follow the evolution of 76 isotopes up to the Iron peak, but we only present here results for the elements having well--measured gradients: C, N, O, Ne, Mg, Al, Si, S, Ar and Fe. The calculated current radial distributions of these elements are compared with large samples of data from \ion{H}{ii} regions, B stars, planetary nebulae and open clusters. We also discuss the time evolution of the abundance gradients. Good fits are achieved with both infall compositions, closer to the observational data in the case of primordial infall. We show that contributions from intermediate--mass stars allow to reproduce the measured abundances of C and N, but the profiles of their ratios to oxygen leave still open the nucleosynthetic origin of these elements. In agreement with several previous works, our models predict a flattening with time of the abundance gradients, most of which taking place during the early galactic evolution.
}

\maketitle

\section{Introduction \label{intro}}
Studies of the chemical evolution of the Galaxy have benefit in recent years from a growing number of high quality observations concerning chemical abundances in stars and in the interstellar medium (see Smartt \cite{smar01a} for a review), from which important constraints on the ingredients of theoretical models can be inferred. In addition, on the theoretical side, new metallicity dependent yields derived from up to date stellar evolutionary calculations covering the whole range of stellar masses have been provided by several groups  (van den Hoek \& Groenengen \cite{vand97}; Forestini \& Charbonnel \cite{fore97}; Marigo et al. \cite{mari96}; Marigo \cite{mari01}, for low and intermediate--mass stars; Woosley \& Weaver \cite{woos95}; Limongi et al. \cite{limo00}, for massive stars).

Among the different observables, the stellar metallicity distribution in the solar neighborhood (G--dwarf metallicity distribution) and the abundance gradients along the galactic disk play a major role. Standard closed--box models of the chemical evolution of the galactic disk cannot adequately reproduce the G--dwarf metallicity distribution, as they produced too many stars at low metallicities. However, the so called infall models, in which the Galaxy is assembled by continuous infall of primordial or lightly enriched gas, are able to reasonably reproduce most of the observational data in the solar ring (Boissier \& Prantzos \cite{bois99}; Chang et al. \cite{chan99}; Portinari \& Chiosi \cite{port99}, that included radial gas flows and the effects of a galactic bar in Portinari \& Chiosi \cite{port00}; Hou et al. {\cite{hou00}; Chiappini et al. \cite{chia01}). Although  different authors use different prescriptions for the basic ingredients (essentially the law for the star formation rate (SFR), the slope and the lower and upper mass limits of the initial mass function (IMF), the gas flows in and out of the Galaxy, and the nucleosynthetic yields) most of them share several basic features. In particular, they usually consider an inside--out scenario for the formation of the galactic disk (Larson \cite{Larson76}) through radially dependent timescales for both the infall and the SFR, IMF constant in space and time, and negligible galactic winds.    

However, the convergence reached by  standard chemical models on the main observables in the solar neighborhood contrast with the results on the abundance gradients in the galactic disk. The existence of such gradients, with single slope values $d\log(X/H)/dr\sim -0.06$ dex kpc$^{-1}$ for most of the elements measured, seems today firmly established from observations of \ion{H}{ii} regions, young O and B stars, planetary nebulae (PN), and open clusters. Several simple models (not including radial flows) are able to reproduce the presently observed trend of the abundance distribution as a function of the galactocentric distance. Nevertheless, they substantially disagree regarding the time evolution of the abundance gradients. Models by Moll\'a et al. (\cite{molla97}), Allen et al. (\cite{allen98}), Boissier \& Prantzos (\cite{bois99}), Portinari \& Chiosi (\cite{port99}) or Hou et al. (\cite{hou00}) obtain steeper gradients in the past, while the opposite is found in papers by Tosi (\cite{tosi88}) or Chiappini et al. (\cite{chia97}). This discrepancy results from the different radial dependence of the enrichment (SFR) and dilution (infall and mass loss from  low--mass stars) processes of the interstellar medium adopted in the models.

Unfortunately, the present observational data do not allow to discriminate between these opposite theoretical results on the history of the galactic abundance gradients. In principle, measures of elemental abundances in PN could help to solve the discrepancy because different types of PN have stellar progenitors of different masses and ages. In the same direction, studies of abundances in open clusters, whose ages cover a quite wide interval, would provide clues to discriminate the history of the galactic gradients. There is some evidence, derived from studies of PN, for a steepening of the gradients in the last few Gyr (Maciel \& Quireza \cite{maci99}), but the uncertainties that still plague the ages attributed to those objects, the lack of good samples in the outer Galaxy, and the difficulty to asses the importance  of dynamical effects on the distribution of old galactic objects prevent us to extract any firm conclusion about the evolution of abundance gradients in the Galaxy from PN studies. On the other hand, iron abundances derived from open clusters of different ages point towards a fairly constant iron gradient for most of the Galaxy lifetime, or even to a flattening of the iron profile during the last Gyrs.  

In this paper we address the chemical evolution in the Milky Way by extending our model for the solar neighborhood (Alib\'es, Labay \& Canal \cite{alib01}, hereafter ALC\cite{alib01}) to the whole galactic halo and disk. We use a standard open model (radial flows are not considered), which treats the formation of the Galaxy by infall of external gas following the double infall scenario of Chiappini et al. (\cite{chia97}). Two different kinds of composition of the infall material are used: primordial and enriched. The enriched one is based on observations of High Velocity Clouds falling towards the galactic disk. Recent measurements of Wakker et al. (\cite{wakk99}) of a massive ($\sim 10^7\, M_{\sun}$) cloud falling into the disk of the Milky Way show that such objects are slightly metal--enriched ($\sim 0.09\, Z_{\sun}$). In ALC\cite{alib01} we showed that we can successfully reproduce the main observational constraints regarding the chemical evolution in the solar neighborhood: G--dwarf metallicity distribution, age--metallicity relation, current values of the star formation and the supernovae rates, the gas and star surface density and the evolution of the main elements up to the Iron peak. As in our previous work on the solar ring, here we calculate the evolution of chemical abundances along the galactic disk for the 76 stable isotopes of the elements comprised between hydrogen and iron, but we will just show here the results obtained for those elements which have been extensively observed in the Milky Way.

The plan of the paper is as follows. In Sect. \ref{model} we briefly remind the main ingredients of our model. In Sect. \ref{data} we present a short review of the currently available observational data on the elemental abundance distributions in the Galaxy. Sect. \ref{results} compares with observations our results for the two assumed compositions of the accreted matter. In Sect. \ref{profiles} and Sect. \ref{gradients} we examine in detail the galactic radial profiles and compare our results with the most recent data on the galactic gradients derived from observations both of young and old objects. In Sect. \ref{gradevol} the time evolution of the abundance gradients is discussed and compared with data from old objects. Finally, in Sect. \ref{conclusions} we summarize our main conclusions.

\section{Overview of the chemical evolution model \label{model}}
As usual, we model the galactic disk by a succession of 1 kpc wide concentric rings, extending from 2 to 20 kpc. Each ring evolves independently, increasing its mass by infall of extragalactic material with primordial or slightly enriched chemical composition. No outflows nor radial flows are considered. Some combinations of infall, SFR and drift velocity pattern of radial flows can successfully reproduce the observed radial profiles in the galactic disk (Portinari \& Chiosi \cite{port00}). Given our poor knowledge of the processes that induce those radial flows (viscosity of the gas layer, differences in angular momentum between the infalling gas and that in the gaseous disk, interactions with spiral density waves), their inclusion in our chemical models would increase the number of adjustable parameters. On the other hand, radial flows tend to steepen the chemical gradients, and the observed values set limitations to any radial mixing. Therefore, we prefer to remain within the frame of static models. We calculate the chemical evolution of each ring by solving the standard integro--differential equations of chemical evolution (Tinsley \cite{tins80}), relaxing the instantaneous recycling approximation. We use the same prescriptions for the infall law, the star formation rate (SFR), the initial mass function (IMF) and the chemical yields as in ALC\cite{alib01}. In the next subsections we briefly describe them, mainly highlighting their radial--dependence characteristics.

\subsection{Infall}
Infall models of galactic chemical evolution ordinarily adopt a single exponentially decreasing law to describe the assembling of the galactic disk. Thus, this kind of models does not take into account the formation and evolution of the halo and thick disk components of the Galaxy, although the final results concerning the properties of the galactic disk only show small differences as compared with those obtained by more complex infall laws. In any case, for consistence with our previous work, we also adopt here the two--infall model developed by Chiappini et al. (\cite{chia97}), that assumes that the Galaxy builds up through two successive accretion episodes. The first one is responsible for the rapid formation of the halo and thick disk, and is followed by a separate, second and more extended episode which assembles the thin disk on timescales that are radially dependent, within the framework of an inside--out scenario for the formation of spiral disks. Hence, we adopt the following infall law:  

\begin{equation}
\frac{d\sigma (r,t)}{dt}=A(r)e^{-t/\tau _{\mathrm{T}}}+B(r)e^{-(t-t_{\mathrm{max}})/\tau _{\mathrm{D}}(r)}
\label{2infall}
\end{equation}

\noindent where $\sigma (r,t)$ is the total surface mass  density, $\tau_\mathrm{T}$ is the timescale for the halo--thick disk phase, taken as 1 Gyr for all $r$, and $\tau_\mathrm{D}(r)$ is the corresponding timescale for the thin disk phase, that we assume to be a linear function of the galactocentric distance, according to the equation 

\begin{equation}
\tau_\mathrm{D}(r)=0.923r-0.846\ \mathrm{Gyr}
\label{taud}
\end{equation}

\noindent obtained by adopting an inner disk value of $\tau_{\mathrm{D}}(r=2$ kpc) $=1$ Gyr, and requiring that in the solar ring ($r_{\sun}=8.5$ kpc) $\tau_\mathrm{D}$ takes the value of 7 Gyr that gave the best results in ALC\cite{alib01}. Finally, $t_{\mathrm{max}}$, the time of maximum accretion into the thin disk, is set as 1 Gyr.

The coefficients $A(r)$ and $B(r)$ in Eq. \ref{2infall} are fixed by imposing that the current total surface mass densities of the thick and thin disks are well reproduced by the model:
\[
A(r)=\frac{\sigma _{\mathrm{T}}(r,t_{\mathrm{G}})}{\tau _{\mathrm{T}}(1-e^{-t_{\mathrm{G}}/\tau _{\mathrm{T}}})}\]
 
\[
B(r)=\frac{[\sigma (r,t_{\mathrm{G}})-\sigma _{\mathrm{T}}(r,t_{\mathrm{G}})]}{\tau _{\mathrm{D}}(r)[1-e^{-(t_{\mathrm{G}}-t_{\mathrm{max}})/\tau _{\mathrm{D}}(r)}]}\]

\noindent where the current total surface density, $\sigma (r,t_{\mathrm{G}})$, and that of the thick disk, $\sigma _{\mathrm{T}}(r,t_{\mathrm{G}})$, are exponentially decreasing functions of the galactocentric distance with the same  characteristic scale length of $r_{\mathrm{D}} = 3.0$ kpc, and the present solar values are 54 and $10\, M_{\sun}$ pc$^{-2}$, respectively. We set the Galaxy age as $t_{\mathrm{G}} = 13$ Gyr.

As stated in the Introduction, we have calculated models for two chemical compositions of the infalling material. In the first case, which we will term in the following ``primordial" model, we consider accretion of primordial material during the whole lifetime of the Galaxy, while a second set of calculations (our ``enriched" model) starts incorporating primordial matter during the quick phase of the halo--thick disk formation and then, in the thin disk epoch, they accrete material with a low ($Z = 0.1\, Z_{\sun}$) but nonzero metallicity, that being inspired by the observations by Wakker et al. (\cite{wakk99}) of high velocity clouds of low metallicity ($Z \sim 0.09\, Z_{\sun}$) falling onto the galactic disk. As shown in ALC\cite{alib01}, these two types of infall models produce comparable results for the characteristics of the solar neighborhood but, as we will see, they exhibit noticeable differences in the outer galaxy where, due to the long infall timescales, the star formation and the corresponding metal enrichment of the interstellar medium have been quite low at all epochs, and metals in the incorporated material tend to level off the radial abundance profiles.

\subsection{Star formation rate}

We adopt the same formulation for the SFR than in ALC\cite{alib01}. The global behaviour of the SFR depends on the surface gas density, $\sigma_{\mathrm{g}}$, according to the Schmidt (\cite{schm59}) law, $\Psi \propto \sigma_{\mathrm{g}} ^m$, with an efficiency coefficient that depends on the galactocentric distance through some power $n$ of the local total mass surface density $\sigma(r,t)$

\begin{equation}
\Psi (r,t)=\nu \frac{\sigma ^{n}(r,t)\sigma _{\mathrm{g}}^{m}(r,t)}{\sigma
^{n+m-1}(r_{\sun },t)}\ \mathrm{M_{\sun } \ pc^{-2} \ Gyr^{-1}}
\label{sfrDR}
\end{equation}

\noindent where $\sigma(r_{\sun },t)$ is introduced as a normalization factor in order to express the efficiency constant $\nu $ in Gyr$^{-1}$. As in ALC\cite{alib01}, we adopt a value of $\nu = 1.2$ which is the one that best reproduce the characteristics of the solar neighborhood. We take $m = 5/3$ and $n = 1/3$ since, according to Dopita and Ryder (\cite{dopi94}), those values fit well the empirical link found by these authors in spiral disks between the H$_\alpha$ emission, tracing current star formation, and the $I$--band surface brightness, a measure of the contribution of the old stellar component and, therefore, of the total surface mass density. We note that recent observations by Kennicutt (\cite{kenn98}) of the correlation between average SFR and surface gas densities in spiral galaxies point towards an exponent in the Schmidt law of $m \sim 1.5$. It is also worth to mention that, as can be seen in Fig. \ref{sfr-gas}, when this SFR law is applied to the observed current gas (Dame \cite{dame93}) and total mass (Rana \cite{rana91}) density profiles, it precisely follows the observed present SFR distribution across the galactic disk (Lyne et al. \cite{lyne85}; Rana \cite{rana91}).

An alternative treatment of the radial dependence of the SFR can be found in Boissier \& Prantzos (\cite{bois99}) and Hou et al. (\cite{hou00}), based on the theory of star formation induced by density waves in spiral galaxies (Wyse \& Silk \cite{wyse89}). Although we will only show here results obtained with the Dopita \& Ryder law, we get very similar conclusions from calculations that use the Wyse \& Silk SFR formulation.

\begin{figure}
\resizebox{\hsize}{!}{\includegraphics{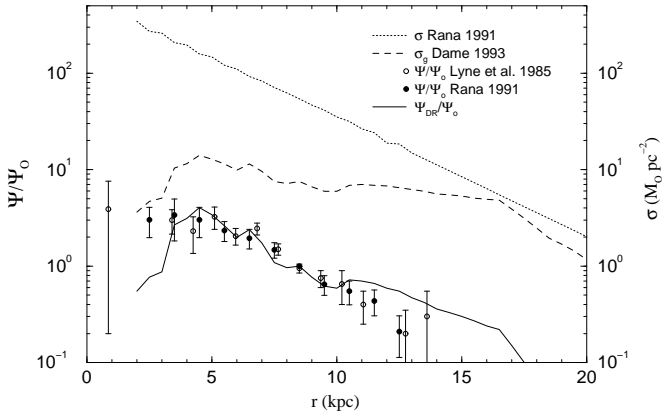}}
\caption{Current surface density profiles of total and gas mass, and the theoretical prescription for the SFR by Dopita \& Ryder (\cite{dopi94}) (normalized to the solar value) applied to these distributions, compared to observational estimates from Lyne et al. (\cite{lyne85}) and Rana (\cite{rana91})}
\label{sfr-gas}
\end{figure}

\subsection{Initial Mass Function}

Despite some claims of variations of the IMF in time and/or in place (see Eisenhauer \cite{eise01}), there are not yet solid enough reasons to abandon the common view of a universal IMF. Moreover, simple models of chemical evolution can only deal with average trends, and in this case constant IMFs are favoured, as noted by Chiappini et al. (\cite{chia00}). Same as in ALC\cite{alib01}, in this work we present results obtained using the IMF from Kroupa et al. (\cite{krou93}), which consists of a three--slope power law. Compared with the classical IMF of Salpeter (\cite{salp55}), it is quite flat for the lowest masses ($x = 0.3$ for $M < 0.5\, M_{\sun}$), it steepens to $x = 1.2$ for $0.5\, M_{\sun} < M < 1\, M_{\sun}$, and for higher masses it has a $x = 1.7$ slope. We adopt a minimum stellar mass of $0.8\, M_{\sun}$ and a maximum of $100\, M_{\sun}$.

We have also calculated models that incorporate the IMF of Scalo (\cite{scal86}). The basic trends of our results are almost unchanged, except for the fact that models that use the IMF for Scalo give slightly steeper abundance gradients, but the differences are so minute that we will not discuss them further in this paper.

\subsection{Yields}

The contribution to the chemical enrichment of the interstellar medium by stars of different masses and metallicities, i.e. the stellar yields, is one of the most important ingredients of models of galactic chemical evolution. In the last years several groups have provided new yields for the complete range of stellar masses and for several metallicities ranging from $0$ to solar or even supersolar. The stellar yields adopted in this paper are the same as in ALC\cite{alib01}. We briefly outline them in the following.

Yields for low-- and intermediate--mass stars are from van den Hoek \& Groenewegen (\cite{vand97}, hereafter HG\cite{vand97}). For massive stars we use the yields from Woosley \& Weaver (\cite{woos95}; hereafter WW\cite{woos95})\footnote{We remind here that, as in our previous solar neighborhood model (ALC\cite{alib01}), we take only half of the iron yields calculated by these authors.}. Type Ia supernovae are included according to the yields from Thielemann et al. ({\cite{thie93}). We also include nova outburst nucleosynthesis (an important source of \element[][13]{C}, \element[][15]{N} and \element[][17]{O}) through the work of Jos\'e \& Hernanz (\cite{jose98}). We note that, despite the efforts of the different groups working in this field to produce accurate stellar yields, all the currently available ones suffer still for severe uncertainties, linked to the poor knowledge of phenomena such as mass loss, rotation, convection, the reaction rate for \element[][12]{C}($\alpha$,$\gamma$)\element[][16]{O}, modeling of core collapse supernovae, etc. 

We also want to mention that, in the inner galaxy, high metallicities are reached and there are no published yields for stars of such high metal contents. The maximum values reached in the sources used here are $Z_{\sun}$ for the WW\cite{woos95} yields, and $2\, Z_{\sun}$ in the case of those of HG\cite{vand97}. We continue to use these yields even when a galactic zone reaches higher metallicities, since we think that extrapolation of the stellar yields beyond these maximum values is a less reliable procedure.

\section{Observational data \label{data}}

We concisely review in this section the main observables for the disk of the Milky Way, relevant to the chemical evolution, which we will take as for reference to compare our results with. As mentioned, for instance, by Boissier \& Prantzos (\cite{bois99}), all the existing observational data only inform us on the present properties of the galactic disk, but not on its past history, excepting the temptative indications obtained from the study of abundances in PNII and PNIII (see below) or in open clusters.

\subsection{Gaseous, stellar and SFR radial profiles}

The current surface gas density distribution is taken from Dame (\cite{dame93}), who gives the observed surface profiles of atomic and molecular hydrogen in the Galaxy. The total gaseous profile is the sum of both, increased by 40\% to take into account helium and heavier elements, and is shown by the dashed line in Fig. \ref{sfr-gas}. As is widely known, its most prominent feature is the molecular ring about 4--5 kpc from the galactic center.

The observed stellar density profile is exponentially decreasing outwards. Recent estimates of the characteristic scale length give values in the range $2.5 - 3$ kpc (Sackett \cite{sack97}; Freudenreich \cite{freu98}). The stellar surface density at the solar ring is $\sigma_{*}(r_{\sun},t_{\mathrm{G}})=35\pm5\, M_{\sun}$ pc$^{-1}$ (Gilmore et al. \cite{gilm89}).

The SFR along the disk of the Milky Way, in terms of its local value, has been estimated from observations of the $H_\alpha$ emissivity (G\"usten \& Mezger \cite{gust82}), the distribution of supernova remnants (Guibert et al. \cite{gui78}), that of molecular gas (Rana \cite{rana91}), or from the surface density of pulsars (Lyne et al. \cite{lyne85}), all of those methods giving results in good agreement with each other. We display in Fig. \ref{sfr-gas} and \ref{gas} the data from the last two references.

\subsection{Abundance gradients} 

Observational studies on the radial profiles of the abundances of several chemical elements have increased during the last decade. Most of them are related to young objects like \ion{H}{ii} regions and B--type stars, with ages clearly inferior to 1 Gyr. Thus, these objects only give information on the present day abundances in the galactic disk. Observations of PN and open clusters could, in principle, give us some insights on the past evolution of the abundance gradients. In Table \ref{hiibpn} we directly collect most of the data available in the literature on the radial gradients in the galactic disk derived from observations of \ion{H}{ii} regions, B stars and PN. Observed iron gradients are summarized in Table \ref{oclu}.

\begin{table*}
\caption{Calculated abundance gradients in the galactic disk, in dex kpc$^{-1}$, and observed values obtained from \ion{H}{ii} regions, B stars and planetary nebulae}
{\centering\begin{tabular}{lcccccccccc}
\hline 
&
\( \Delta r \)&C&N&O&Ne&Mg&Al&Si&S&Ar\\
\hline 
\hline 
\textbf{\footnotesize \ion{H}{ii}}
{\footnotesize }&{\footnotesize }&{\footnotesize }&
{\footnotesize }&{\footnotesize }&{\footnotesize }&{\footnotesize }&
{\footnotesize }&{\footnotesize }&{\footnotesize }&{\footnotesize }\\
\hline 
{\footnotesize Shaver et a. \cite{shav83}}&
{\footnotesize 5-14}&{\footnotesize }&{\footnotesize $-0.090$}&
{\footnotesize $-0.070$}&{\footnotesize }&{\footnotesize }&{\footnotesize }&
{\footnotesize }&{\footnotesize $-0.010$}&{\footnotesize $-0.060$}\\
{\footnotesize }&
{\footnotesize }&{\footnotesize }&{\footnotesize \( \pm 0.015 \)}&
{\footnotesize \( \pm 0.015 \)}&{\footnotesize }&{\footnotesize }&{\footnotesize }&
{\footnotesize }&{\footnotesize \( \pm 0.020 \)}&{\footnotesize \( \pm 0.015 \)}\\
{\footnotesize Simpson et al. \cite{simp95}}&
{\footnotesize 0-10}&{\footnotesize }&{\footnotesize $-0.100$}&{\footnotesize }&{\footnotesize $-0.080$}&{\footnotesize }&{\footnotesize }&{\footnotesize }&{\footnotesize $-0.070$}&{\footnotesize }\\
{\footnotesize }&{\footnotesize }&{\footnotesize }&
{\footnotesize \( \pm 0.020\)}&{\footnotesize }&{\footnotesize \( \pm 0.020 \)}&
{\footnotesize }&{\footnotesize }&{\footnotesize }&{\footnotesize \( \pm 0.020 \)}&{\footnotesize }\\
{\footnotesize V\'{\i}lchez \& Esteban \cite{vilc96}}&
{\footnotesize 12-18}&{\footnotesize }&
{\footnotesize $-0.009$}&{\footnotesize $-0.036$}&{\footnotesize }&{\footnotesize }&
{\footnotesize }&{\footnotesize }&{\footnotesize $-0.041$}&{\footnotesize }\\
{\footnotesize }&{\footnotesize }&{\footnotesize }&{\footnotesize \( \pm 0.020 \)}&
{\footnotesize \( \pm 0.020 \)}&{\footnotesize }&{\footnotesize }&
{\footnotesize }&{\footnotesize }&{\footnotesize \( \pm 0.020 \)}&
{\footnotesize }\\
{\footnotesize Afflerbach et al. \cite{affl97}}&{\footnotesize 0-12}&{\footnotesize }&
{\footnotesize $-0.072$}&{\footnotesize $-0.064$}&{\footnotesize }&{\footnotesize }&
{\footnotesize }&{\footnotesize }&{\footnotesize $-0.063$}&{\footnotesize }\\
{\footnotesize }&{\footnotesize }&{\footnotesize }&{\footnotesize \( \pm 0.006 \)}&
{\footnotesize \( \pm 0.009 \)}&{\footnotesize }&{\footnotesize }&{\footnotesize }&
{\footnotesize }&{\footnotesize \( \pm 0.006 \)}&{\footnotesize }\\
{\footnotesize Rudolph et al. \cite{rudo97}}&
{\footnotesize 0-17}&{\footnotesize }&{\footnotesize $-0.111$}&{\footnotesize }&
{\footnotesize }&{\footnotesize }&{\footnotesize }&{\footnotesize }&
{\footnotesize $-0.079$}&{\footnotesize }\\
{\footnotesize }&{\footnotesize }&{\footnotesize }&
{\footnotesize \( \pm 0.012 \)}&{\footnotesize }&{\footnotesize }&
{\footnotesize }&{\footnotesize }&{\footnotesize }&{\footnotesize \( \pm 0.009 \)}&{\footnotesize }\\
{\footnotesize Esteban et al. \cite{este99}}&
{\footnotesize 6-9}&{\footnotesize $-0.133$}&{\footnotesize $-0.048$}&
{\footnotesize $-0.049$}&{\footnotesize $-0.045$}&{\footnotesize}&{\footnotesize}&
{\footnotesize}&{\footnotesize $-0.055$}&{\footnotesize$-0.044$}\\
{\footnotesize }&{\footnotesize }&{\footnotesize \( \pm 0.022 \)}&
{\footnotesize \( \pm 0.017 \)}&{\footnotesize \( \pm 0.017 \)}&
{\footnotesize \( \pm 0.017 \)}&{\footnotesize }&{\footnotesize }&{\footnotesize }&
{\footnotesize \( \pm 0.017 \)}&{\footnotesize \( \pm 0.030 \)}\\
{\footnotesize Deharveng et al. \cite{deha00}}&
{\footnotesize 6-18}&{\footnotesize }&{\footnotesize }&{\footnotesize $-0.039$}&
{\footnotesize }&{\footnotesize }&{\footnotesize }&{\footnotesize }&
{\footnotesize }&{\footnotesize }\\
{\footnotesize }&{\footnotesize }&{\footnotesize }&{\footnotesize }&
{\footnotesize \( \pm 0.005 \)}&{\footnotesize }&{\footnotesize }&
{\footnotesize }&{\footnotesize }&{\footnotesize }&{\footnotesize }\\
\hline 
\textbf{\footnotesize B stars}{\footnotesize }&
{\footnotesize }&{\footnotesize }&{\footnotesize }&{\footnotesize }&
{\footnotesize }&{\footnotesize }&{\footnotesize }&{\footnotesize }&
{\footnotesize }&{\footnotesize }\\
\hline 
{\footnotesize Gehren et al. \cite{gehr85}}&
{\footnotesize 8-18}&{\footnotesize }&{\footnotesize }&{\footnotesize $-0.010$}&
{\footnotesize }&{\footnotesize }&{\footnotesize }&{\footnotesize }&
{\footnotesize }&{\footnotesize }\\
{\footnotesize }&{\footnotesize }&{\footnotesize }&{\footnotesize }&
{\footnotesize \( \pm 0.020 \)}&{\footnotesize }&{\footnotesize }&
{\footnotesize }&{\footnotesize }&{\footnotesize }&{\footnotesize }\\
{\footnotesize Fitzsimmons et al. \cite{fitz92}}&
{\footnotesize 6-13}&{\footnotesize }&{\footnotesize }&{\footnotesize $-0.030$}&
{\footnotesize }&{\footnotesize }&{\footnotesize }&{\footnotesize }&
{\footnotesize }&{\footnotesize }\\
{\footnotesize }&{\footnotesize }&{\footnotesize }&{\footnotesize }&
{\footnotesize \( \pm 0.020 \)}&{\footnotesize }&{\footnotesize }&
{\footnotesize }&{\footnotesize }&{\footnotesize }&{\footnotesize }\\
{\footnotesize Kaufer et al. \cite{kauf94}}&
{\footnotesize 6-17}&{\footnotesize }&{\footnotesize $-0.026$}&
{\footnotesize $-0.000$}&{\footnotesize }&{\footnotesize }&{\footnotesize }&
{\footnotesize }&{\footnotesize }&{\footnotesize }\\
{\footnotesize }&{\footnotesize }&{\footnotesize }&{\footnotesize \( \pm 0.009 \)}&
{\footnotesize \( \pm 0.009 \)}&{\footnotesize }&{\footnotesize }&
{\footnotesize }&{\footnotesize }&{\footnotesize }& {\footnotesize }\\
{\footnotesize Kilian-M. et al. \cite{kili94}}&
{\footnotesize 6-15}&{\footnotesize $-0.001$}&{\footnotesize $-0.017$}&
{\footnotesize $-0.021$}&{\footnotesize $-0.043$}&{\footnotesize $-0.020$}&
{\footnotesize $-0.002$}&{\footnotesize$+0.000$}&{\footnotesize $-0.026$}&
{\footnotesize }\\
{\footnotesize }&{\footnotesize }&{\footnotesize \( \pm 0.015 \)}&
{\footnotesize \( \pm 0.020 \)}&{\footnotesize \( \pm 0.012 \)}&
{\footnotesize \( \pm 0.011 \)}&{\footnotesize \( \pm 0.011 \)}&
{\footnotesize \( \pm 0.019 \)}&{\footnotesize \( \pm 0.018 \)}&
{\footnotesize \( \pm 0.025 \)}&
{\footnotesize }\\
{\footnotesize Smartt \& Rolleston \cite{smar97}}&
{\footnotesize 6-18}&{\footnotesize }&{\footnotesize }&{\footnotesize $-0.070$}&
{\footnotesize }&{\footnotesize }&{\footnotesize }&{\footnotesize }&
{\footnotesize }&{\footnotesize }\\
{\footnotesize }&{\footnotesize }&{\footnotesize }&{\footnotesize }&
{\footnotesize \( \pm 0.010 \)}&{\footnotesize }&{\footnotesize }&
{\footnotesize }&{\footnotesize }&{\footnotesize }&{\footnotesize }\\
{\footnotesize Gummersbach et al. \cite{gumm98}}&
{\footnotesize 5-14}&{\footnotesize $-0.035$}&{\footnotesize $-0.078$}&
{\footnotesize $-0.067$}&{\footnotesize }&{\footnotesize $-0.082$}&
{\footnotesize $-0.045$}&{\footnotesize $-0.107$}&{\footnotesize }&
{\footnotesize }\\
{\footnotesize }&{\footnotesize }&{\footnotesize \( \pm 0.014 \)}&
{\footnotesize \( \pm 0.023 \)}&{\footnotesize \( \pm 0.024 \)}&
{\footnotesize }&{\footnotesize \( \pm 0.026 \)}&
{\footnotesize \( \pm 0.023 \)}&{\footnotesize \( \pm 0.028 \)}&
{\footnotesize }&{\footnotesize }\\
{\footnotesize Rolleston et al. \cite{roll00}}&
{\footnotesize 6-18}&{\footnotesize$-0.070$}&{\footnotesize $-0.090$}&
{\footnotesize $-0.067$}&{\footnotesize }&{\footnotesize $-0.070$}&
{\footnotesize $-0.050$}&{\footnotesize $-0.060$}&{\footnotesize }&
{\footnotesize }\\
{\footnotesize }&{\footnotesize }&{\footnotesize \( \pm 0.020 \)}&
{\footnotesize \( \pm 0.010 \)}&{\footnotesize \( \pm 0.008 \)}&
{\footnotesize }&{\footnotesize \( \pm 0.010 \)}&
{\footnotesize \( \pm 0.015 \)}&{\footnotesize \( \pm 0.010 \)}&
{\footnotesize }&{\footnotesize }\\
{\footnotesize Smartt et al. \cite{smar01b}}&
{\footnotesize 0-18}&{\footnotesize $-0.070$}&{\footnotesize $-0.060$}&
{\footnotesize }&{\footnotesize }&{\footnotesize $-0.090$}&
{\footnotesize $-0.050$}&{\footnotesize $-0.060$}&{\footnotesize }&
{\footnotesize }\\
{\footnotesize }&{\footnotesize }&{\footnotesize \( \pm 0.020 \)}&
{\footnotesize \( \pm 0.020 \)}&{\footnotesize }&{\footnotesize }&
{\footnotesize \( \pm 0.020 \)}&{\footnotesize \( \pm 0.010 \)}&
{\footnotesize \( \pm 0.010 \)}&{\footnotesize }&
{\footnotesize }\\
\hline 
\textbf{\footnotesize PN}{\footnotesize }&
{\footnotesize }&{\footnotesize }&{\footnotesize }&{\footnotesize }&
{\footnotesize }&{\footnotesize }&{\footnotesize }&{\footnotesize }&
{\footnotesize }&{\footnotesize }\\
\hline 
{\footnotesize Pasquali \& Perinotto \cite{pasq93}}&
{\footnotesize 1-16}&{\footnotesize }&{\footnotesize $-0.050$}&
{\footnotesize $-0.030$}&{\footnotesize $-0.050$}&{\footnotesize }&
{\footnotesize }&{\footnotesize }&{\footnotesize }&{\footnotesize }\\
{\footnotesize \hskip 0.5cm} {\footnotesize (PN I - II)}&
{\footnotesize }&{\footnotesize }&{\footnotesize \( \pm 0.010 \)}&
{\footnotesize \( \pm 0.010 \)}&{\footnotesize \( \pm 0.020 \)}&
{\footnotesize }&{\footnotesize }&{\footnotesize }&{\footnotesize }&
{\footnotesize }\\
{\footnotesize Maciel \& K\"oppen \cite{maci94} }&
{\footnotesize 4-13}&{\footnotesize }&{\footnotesize }&{\footnotesize $-0.030$}&
{\footnotesize $-0.004$}&{\footnotesize }&{\footnotesize }&{\footnotesize }&
{\footnotesize $-0.075$}&{\footnotesize $-0.060$}\\
{\footnotesize \hskip 0.5cm} {\footnotesize (PN I)}&
{\footnotesize }&{\footnotesize }&{\footnotesize }&
{\footnotesize \( \pm 0.007 \)}&{\footnotesize \( \pm 0.008 \)}&
{\footnotesize }&{\footnotesize }&{\footnotesize }&
{\footnotesize \( \pm 0.008 \)}&{\footnotesize \( \pm 0.008 \)}\\
{\footnotesize Maciel \& K\"oppen \cite{maci94}}&
{\footnotesize 4-13}&{\footnotesize }&{\footnotesize }&
{\footnotesize $-0.069$}&{\footnotesize $-0.056$}&
{\footnotesize }&{\footnotesize }&{\footnotesize }&
{\footnotesize $-0.067$}&{\footnotesize $-0.051$}\\
{\footnotesize \hskip 0.5cm} {\footnotesize (PN II)}&
{\footnotesize }&{\footnotesize }&{\footnotesize }&
{\footnotesize \( \pm 0.006 \)}&{\footnotesize \( \pm 0.007 \)}&
{\footnotesize }&{\footnotesize }&{\footnotesize }&
{\footnotesize \( \pm 0.006 \)}&{\footnotesize \( \pm 0.006 \)}\\
{\footnotesize Maciel \& K\"oppen \cite{maci94}}&
{\footnotesize 4-13}&{\footnotesize }&{\footnotesize }&
{\footnotesize $-0.058$}&{\footnotesize $-0.041$}&
{\footnotesize }&{\footnotesize }&{\footnotesize }&
{\footnotesize $-0.063$}&{\footnotesize $-0.034$}\\
{\footnotesize \hskip 0.5cm} {\footnotesize (PN III)}&
{\footnotesize }&{\footnotesize }&{\footnotesize }&
{\footnotesize \( \pm 0.008 \)}&{\footnotesize \( \pm 0.008 \)}&
{\footnotesize }&{\footnotesize }&{\footnotesize }&
{\footnotesize \( \pm 0.010 \)}&{\footnotesize \( \pm 0.010 \)}\\
{\footnotesize Maciel \& Quireza \cite{maci99}}&
{\footnotesize 3-14}&{\footnotesize }&{\footnotesize }&
{\footnotesize $-0.058$}&{\footnotesize $-0.036$}&
{\footnotesize }&{\footnotesize }&{\footnotesize }&
{\footnotesize $-0.077$}&{\footnotesize $-0.051$}\\
{\footnotesize \hskip 0.5cm} {\footnotesize (PN II)}&
{\footnotesize }&{\footnotesize }&{\footnotesize }&
{\footnotesize \( \pm 0.007 \)}&{\footnotesize \( \pm 0.010 \)}&
{\footnotesize }&{\footnotesize }&{\footnotesize }&
{\footnotesize \( \pm 0.011 \)}&{\footnotesize \( \pm 0.010 \)}\\
{\footnotesize Martins \& Veigas \cite{mart00}}&
{\footnotesize 5-13}&{\footnotesize }&{\footnotesize $-0.084$}&
{\footnotesize $-0.054$}&{\footnotesize$-0.069$}&{\footnotesize }&
{\footnotesize }&{\footnotesize }&{\footnotesize $-0.064$}&
{\footnotesize }\\
{\footnotesize \hskip 0.5cm} {\footnotesize (PN II)}&
{\footnotesize }&{\footnotesize }&{\footnotesize \( \pm 0.034 \)}&
{\footnotesize \( \pm 0.019 \)}&{\footnotesize \( \pm 0.034 \)}&
{\footnotesize }&{\footnotesize }&{\footnotesize }&
{\footnotesize \( \pm 0.035 \)}&
\\
\hline
\textbf{\footnotesize Model results}{\footnotesize }&
{\footnotesize }&{\footnotesize }&{\footnotesize }&{\footnotesize }&
{\footnotesize }&{\footnotesize }&{\footnotesize }&{\footnotesize }&
{\footnotesize }&{\footnotesize }\\
\hline 
{\footnotesize Enriched model}&
{\footnotesize 4-16}&{\footnotesize $-0.051$}&{\footnotesize $-0.066$}&
{\footnotesize $-0.047$}&{\footnotesize $-0.043$}&
{\footnotesize $-0.052$}&{\footnotesize $-0.061$}&{\footnotesize $-0.052$}&
{\footnotesize $-0.054$}&{\footnotesize $-0.057$}\\
{\footnotesize }&
{\footnotesize 4-10}&{\footnotesize $-0.045$}&{\footnotesize $-0.054$}&
{\footnotesize $-0.040$}&{\footnotesize $-0.035$}&
{\footnotesize $-0.040$}&{\footnotesize $-0.041$}&{\footnotesize $-0.039$}&
{\footnotesize $-0.040$}&{\footnotesize $-0.039$}\\
{\footnotesize }&
{\footnotesize 10-16}&{\footnotesize $-0.064$}&{\footnotesize $-0.085$}&
{\footnotesize $-0.062$}&{\footnotesize $-0.058$}&
{\footnotesize $-0.070$}&{\footnotesize $-0.089$}&{\footnotesize $-0.072$}&
{\footnotesize $-0.076$}&{\footnotesize $-0.082$}\\
\hline
{\footnotesize Primordial model}&
{\footnotesize 4-16}&{\footnotesize $-0.058$}&{\footnotesize $-0.074$}&
{\footnotesize $-0.053$}&{\footnotesize $-0.048$}&
{\footnotesize $-0.061$}&{\footnotesize $-0.077$}&{\footnotesize $-0.058$}&
{\footnotesize $-0.062$}&{\footnotesize $-0.065$}\\
{\footnotesize }&
{\footnotesize 4-10}&{\footnotesize $-0.049$}&{\footnotesize $-0.058$}&
{\footnotesize $-0.043$}&{\footnotesize $-0.038$}&
{\footnotesize $-0.045$}&{\footnotesize $-0.047$}&{\footnotesize $-0.043$}&
{\footnotesize $-0.045$}&{\footnotesize $-0.044$}\\
{\footnotesize }&
{\footnotesize 10-16}&{\footnotesize $-0.076$}&{\footnotesize $-0.100$}&
{\footnotesize $-0.071$}&{\footnotesize $-0.065$}&
{\footnotesize $-0.085$}&{\footnotesize $-0.117$}&{\footnotesize $-0.082$}&
{\footnotesize $-0.088$}&{\footnotesize $-0.094$}\\
%{\footnotesize primordial model (IMF Scalo)}&
%{\footnotesize 4-16}&{\footnotesize $-0.060$}&{\footnotesize $-0.070$}&
%{\footnotesize $-0.060$}&{\footnotesize $-0.060$}&
%{\footnotesize $-0.068$}&{\footnotesize $-0.079$}&{\footnotesize $-0.059$}&
%{\footnotesize $-0.063$}&{\footnotesize $-0.067$}\\
\hline
\end{tabular}\footnotesize \par}
\label{hiibpn}
\end{table*}

Since the pioneering work by Shaver et al. (\cite{shav83}), the analysis of the emission lines from ionized regions around massive star associations, in the optical and the IR region of the spectrum, have been used to obtain chemical abundance determinations of O, N, S, and, in a lesser extent, of Ar and Ne, in \ion{H}{ii} regions located over a wide range of galactocentric distances (from 0 to 18 kpc). The main uncertainties that affect these measurements are related to the determination of the electronic temperatures (derived by different techniques, varying from direct measurements using suitable diagnostic lines to photoionization models that require a priori assumptions for the flux of ionizing photons), and also to the importance of the temperature fluctuations in the gaseous nebulae, which have not been extensively considered. Most of these studies agree on a galactic gradient of O and S of $\sim -0.07$ dex kpc$^{-1}$, and a little bit steeper for N. Some authors (V\'{\i}lchez \& Esteban \cite{vilc96}) suggest a flattening of the gradients in the outer Galaxy, but the most recent analyses favour a unique slope (Afflerbach et al. \cite{affl97}; Rudolph et al. \cite{rudo97}). However, this general concordance has recently being challenged by Deharveng et al. (\cite{deha00}). They have determined O abundances in a sample of \ion{H}{ii} regions located at galactocentric distances between 6.6 and 18 kpc by means of a double temperature model of \ion{H}{ii} regions. As in most of the previous work, their data do not show any flattening of the O distribution in the external regions, but they obtain an oxygen gradient of $-0.039\pm0.005$ dex kpc$^{-1}$, almost a factor of 2 flatter than the traditionally quoted value but similar to that from Esteban et al. (\cite{este99}). Therefore, even if the existence of abundance gradients derived from \ion{H}{ii} regions is unquestionable, more observational effort is needed to clearly establish their actual magnitude.

During the early 90's, the measurements from \ion{H}{ii} regions showed much steeper gradients than those obtained from the photospheric composition of young B--type stars, which are supposed to reflect the composition of the material from which they formed, except for the rapidly rotating B stars, where there could have been some mixing of atmospheric and core material (Maeder \cite{maed87}). Spectra of those bright stars show optical lines of C, N, O, Mg, Si and, in high--quality spectra, also Al, S, and Fe. A series of studies (Gehren et al. \cite{gehr85}; Fitzsimmons et al. \cite{fitz92}; Kaufer et al. \cite{kauf94}; Kilian-Montenbruck et al. \cite{kili94}) detected no gradients or gradients much flatter than those determined from \ion{H}{ii} regions. Nevertheless, recent analyses of homogeneous samples of B stars in a consistent manner (a crucial factor to obtain reliable measures when using stars) by Smartt \& Rolleston (\cite{smar97}), Gummersbach et al (\cite{gumm98}), Rolleston et al. (\cite{roll00}), and Smartt et al. (\cite{smar01b}) confirmed the existence of gradients similar to those measured in ionized nebulae. Furthermore, the study by Smartt et al. (\cite{smar01b}), which included four stars near the Galactic Center, found little evidences of O abundance variations inside 9 kpc, i.e. a flattening of the O profile in the inner disk, at odds with the rest of the studied elements. 

The analysis of emission lines in planetary nebulae (PN) is another popular method to measure abundances in different regions of the galactic disk. Besides their greater statistical importance owing to the larger samples available as compared with studies of \ion{H}{ii} regions and B--type stars, PN could give, in principle, information about the evolution of the abundance profiles. Following the original classification of PN by Peimbert (\cite{peim78}), revised by Pasquali \& Perinotto (\cite{pasq93}), disk PN are classified into three types according to their chemical and dynamical characteristics. PNI are He and N rich, belong to the thin disk, show peculiar velocities  below $60$ km s$^{-1}$, and are associated with 2.4 -- $8\, M_{\sun}$ progenitors, thus having ages less than 1 Gyr. PNII are also members of the thin disk with low peculiar velocities, but they are not particularly enriched in He nor N; it is supposed that they come from stars in the mass range 1.2 -- $2.4\, M_{\sun}$, which correspond to ages comprised between 1 and 7 Gyr. As for PNIII, located in the thick disk with velocities above $60$ km s$^{-1}$, their assumed progenitors are stars in the mass range 1 -- $1.2\,  M_{\sun}$, these then being the oldest disk PN, with ages in excess of 7 Gyr. However, large uncertainties are involved in the precise determination of distances, ages, and masses of PN. In addition, there exists the possibility that the radial distribution of old disk PN might have been altered by  dynamical effects through diffusion processes along the disk that could affect the gradients determined from those objects. Also, it is not clear whether PN show exactly the metal abundances of the gas at the time of their birth. Some enrichment by the products from the CN cycle affect all types of PN. In the case of PNI, whose progenitors are massive intermediate--mass stars, a possible enrichment by ON cycle products may exist, which could induce misleading oxygen abundance measurements in these nebulae. Recent studies (Pasquali \& Perinotto \cite{pasq93}; Maciel \& K\"oppen \cite{maci94}; Maciel \& Quireza \cite{maci99}) determined gradients of O and S about $-0.06$  dex kpc$^{-1}$ for the available sample of PNII (representative of the thin disk) and a bit flatter, $\sim -0.04$  dex kpc$^{-1}$, for Ne. Some indications of shallower gradients in the outer ($r>10$ kpc) Galaxy exist in those data, as well as a slight tendency of the gradients to flatten with time, but due to the considerable uncertainties noted above none of these points can be taken as definitive.

\begin{table}
\caption{Calculated iron gradients and observations in open clusters}
{\centering \begin{tabular}{lcc}
\hline 
\textbf{Open Clusters}&
\( \Delta r \)&
Fe \\
\hline 
\hline 
{\footnotesize Janes \cite{jane79}}&{\footnotesize 8-14}&{\footnotesize $-0.050\pm0.008$ }\\
{\footnotesize Cameron \cite{came85}}&{\footnotesize 7-11}&{\footnotesize $-0.110\pm 0.020$}\\
{\footnotesize Friel \& Janes \cite{frie93}}&{\footnotesize 8-16}&{\footnotesize$-0.095\pm0.017$}\\
{\footnotesize Piatti et al. \cite{piat95}}&{\footnotesize 7-14}&{\footnotesize $-0.070\pm0.010$}\\
{\footnotesize Twarog et al. \cite{twar97}}&{\footnotesize 6-16}&{\footnotesize $-0.067\pm0.008$}\\
{\footnotesize }&{\footnotesize 6-10}&{\footnotesize $-0.023\pm0.017$}\\
{\footnotesize }&{\footnotesize 10-16}&{\footnotesize $-0.004\pm0.018$}\\
{\footnotesize Carraro et al. \cite{carr98}}&{\footnotesize 7-16}&{\footnotesize $-0.085\pm0.008$}\\
{\footnotesize Friel \cite{frie99}}&{\footnotesize 7-16}&{\footnotesize $-0.060\pm0.010$}\\
\hline 
\textbf{Model results}&&\\
\hline 
{\footnotesize Enriched model}&{\footnotesize 4-16}&{\footnotesize $-0.059$ }\\
{\footnotesize }&{\footnotesize 4-10}&{\footnotesize $-0.055$ }\\
{\footnotesize }&{\footnotesize 10-16}&{\footnotesize $-0.069$ }\\
\hline
{\footnotesize Primordial model}&{\footnotesize 4-16}&{\footnotesize $-0.064$ }\\
{\footnotesize }&{\footnotesize 4-10}&{\footnotesize $-0.059$ }\\
{\footnotesize }&{\footnotesize 10-16}&{\footnotesize $-0.079$ }\\

\hline
\end{tabular}\footnotesize \par}
\label{oclu}
\end{table}

Finally, abundances of Fe and its galactic gradient have been mainly determined by means of spectroscopic and photometric indices of stellar populations in open clusters. Early determinations (Janes \cite{jane79}; Cameron \cite{came85}) found gradients of $-0.06$ to $-0.09$ dex kpc$^{-1}$ over a range of galactocentric distances from 7 to 16 kpc, a trend confirmed by Friel \& Janes (\cite{frie93}). However, Twarog et al. (\cite{twar97}), in an analysis of a sample of 63 clusters spanning a wide interval of ages, did not find any linear trend for the Fe abundances in the galactic disk. These authors suggest that a step function, instead of a unique slope, would better describe the distribution, with the discontinuity located at 10 kpc. Further studies by Carraro et al. (\cite{carr98}) did not shown evidence of any abrupt discontinuity and their data fit a unique slope relation, in agreement with earlier measurements. A similar conclusion was reached by Friel (\cite{frie99}), using high--resolution abundance determinations for metallicity calibrators, who finds a slope of $\sim -0.06\pm 0.01$ dex kpc$^{-1}$, shallower but still compatible with previous values (Table \ref{oclu} summarizes the Fe gradients derived from available data). In any case, our current knowledge on the iron gradient as derived from open clusters is far from being clear. When comparing these results and those corresponding to extreme Pop I objects (\ion{H}{ii} and B--type stars), we must note that open clusters can be as old as 8 Gyr, so that they do not trace the young component of the galactic disk. It is well known that Fe is mainly a product of Type Ia supernovae and thus its abundance history is very different from that of other elements, like oxygen and sulphur, which are almost exclusively synthesized in massive stars. The fact that the slope of the Fe gradient is similar to that of O is therefore surprising, and might be simply coincidental. To make things even more confused, the slope of the Fe gradient in open clusters shows no dependence on age (Friel \cite{frie99}), in other words, its gradient has not evolved significantly during most of the lifetime of the disk. This fact is in clear contradiction with the well--known age--metallicity relation from F stars (Edvardsson et al. \cite{edva93}; Rocha-Pinto et al. \cite{roch00}). When in the next Section we compare the calculated Fe abundances across the disk with measures in open clusters, the reader should remember all the uncertainties affecting these data.

\section{Model results versus observational constraints \label{results}}

We showed in ALC\cite{alib01} that our models nicely reproduce the chemical evolution of the solar neighborhood for both compositions of the accreted matter, although a little better fit was obtained for the enriched composition. In this section we compare our results for the primordial and the enriched infall models with the available observational data on the properties of the Milky Way disk. 

\subsection{Radial profiles: gas, stars and SFR \label{profiles}}

The results obtained for the radial distributions of gas, stars and SFR are displayed in Fig. \ref{gas}. Our model reasonably follows the Dame (\cite{dame93}) data on gas surface density, $\sigma_{\mathrm{g}}$, although, as it usually happens in models that do not consider radial flows nor the dynamical influence of a galactic bar (Portinari \& Chiosi \cite{port99}, \cite{port00}), the peak in gas density corresponding to the molecular ring at 4 kpc is clearly less marked than the observed one. The calculated stellar profile, which is exponentially decreasing, is also in very good correspondence with the observations. We obtain a radial scale length of 2.8 kpc, just in the range of 2.5--3 kpc given by Sackett (\cite{sack97}). Finally, the observed current star formation rate normalized to its solar value ($\Psi/\Psi_{\sun}$) is nicely reproduced by our calculations outside 4--5 kpc, while in the most internal part of the galactic disk, even if still compatible with the observational limits, we obtain values a bit lower than the data.

In summary, the model gives a good representation of the main characteristics of the galactic disk, in particular outside 4 kpc, which corresponds to the regions where most of the data on the abundance gradients have been obtained. Therefore, we can confidently go on with the discussion on the metal abundances across the disk. 

\begin{figure}
\resizebox{\hsize}{!}{\includegraphics{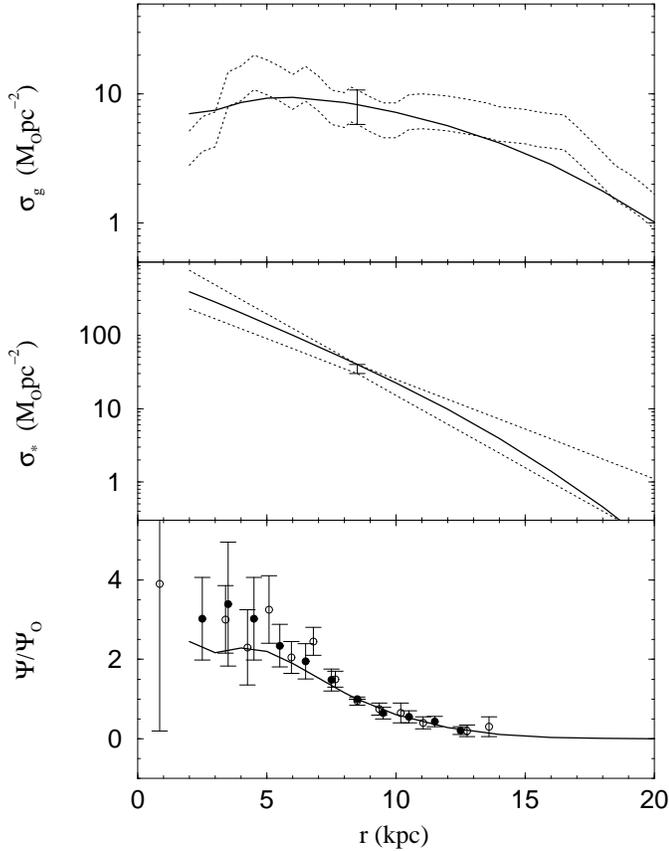}}
\caption{Results for the gas and star surface density profiles, and normalized current star formation rate distribution compared with observations. Data: gas surface density from Dame (\cite{dame93})  accounting for 40\% of He and the observational error; star surface density using $\sigma_*(r_{\sun},t_{\mathrm{G}})=35\pm5\, M_{\sun}$ pc$^{-1}$ (Gilmore et al. \cite{gilm89}) and $r_* = 2.5-3$ kpc (Sackett \cite{sack97}); current SFR from Lyne et al. (\cite{lyne85}), from the distribution of pulsars, and Rana (\cite{rana91}), from the distribution of molecular gas}
\label{gas}
\end{figure}

\subsection{Current abundance gradients and profiles of abundance ratios \label{gradients}}

We begin by discussing the present abundance gradients and the profiles of abundance ratios to oxygen for the chemical elements with measured abundances. Their time evolution will be presented in Sect. \ref{gradevol}. In Fig. \ref{OH-R} to \ref{FeH-R} we show the radial distributions of C, N, O, Ne, Mg, Al, Si, S, Ar and Fe, as well as the profiles of their abundance ratio to O, obtained for both the enriched (solid lines) and the primordial (dashed lines) models, and we compare them with most of the existing data from young objects (\ion{H}{ii} regions, early B--type stars and PNI). In the case of Fe, its abundance has essentially been measured in open clusters, thus data in the Fe graph cover a wide interval of ages. The calculated values of the abundance gradients for all the chemical elements considered appear at the bottom of Table \ref{hiibpn}, except for Fe which are included in Table \ref{oclu}. We give the gradients over an interval of galactocentric distances of 4 to 16 kpc, thus covering both the inner Galaxy (4--10 kpc) and the outer Galaxy (10--16 kpc).

Models of galactic chemical evolution that include a radial dependence of the SFR and the infall rate (the ``inside--out" scenario), like ours, naturally produce more or less steep abundance gradients across the disk. We remind that the parameters of our model were tuned to fit the solar neighborhood constraints, in particular the chemical abundances of the Sun at the moment of its birth 4.5 Gyr ago (see ALC\cite{alib01}). It is well known that the solar neighborhood is more oxygen--rich than the local interstellar medium. This is the reason why the majority of the calculated radial abundance distributions shown in Fig. \ref{OH-R} to \ref{FeH-R} only fit the more metal--rich measurements. Although several suggestions have been put forward to account for the solar overabundances (orbital diffusion, dust depletion, enrichment of the protosolar nebula by neighboring SN II...), there is no convincing in contradiction with predictions from standard chemical models where metallicity is an increasing function of time. However, whatever the real oxygen abundance in the present solar ring may be, what specially counts for our discussion is the radial behaviour of the abundances, i.e. their gradients, since the particular election of the zero--point does not alter the radial characteristics.

We obtain a general good agreement with the data. Results for both compositions of the infalling matter are quite similar in the inner disk, because the high metallicity reached there, due to the high efficiency of star formation, erases any possible influence of a non--primordial composition of the accreted matter. However, in the outer disk, where the SFR remains low during the whole life of the Galaxy, the metallicity is modest even at the present epoch, and models with enriched infall give larger abundances of metals and shallow gradients as compared with those obtained when the accreted matter is assumed to be of primordial composition.

The inner disk abundance profiles are flatter than in the external regions, a common result in most models of galactic chemical evolution (Portinari \& Chiosi \cite{port99}; Boissier \& Prantzos \cite{bois99}; Hou et al. \cite{hou00}). The reason for this flattening has already been discussed in those papers: due to the initially high SFR in the central regions, the metal enrichment is appreciably diluted by the low metallicity gas ejected by the large amount of long--lived, low--mass stars formed at early times from metal--poor gas, an effect that can not be compensated by the ejecta of massive stars at late times, when very little gas is left and the SFR is drastically reduced. This is not the situation in the outer disk, where the surface gas density remains low during the whole evolution and the population of old stars is scarce. It must be noted that the central flattening of the abundance profiles is overestimated by our models. The WW\cite{woos95} stellar yields for massive stars only cover up to solar metallicity, and the HG\cite{vand97} yields for intermediate--mass stars just arrive to twice solar. Then, the use of these yields when metallicities higher than solar are reached in the central zones underestimates the metal enrichment, especially for those elements that have stellar yields that depend on the stellar metal contents (odd--even elements, for instance). 

Since most of the studied elements are produced in massive stars and have stellar yields hardly dependent on $Z$, i.e. they are primary elements, we get flat [X/O] ratios across the galactic disk. The two exceptions are Al, whose yields clearly show the odd--even effect, and N which is mildly secondary in our intermediate--mass stars.
 
In the remaining of this subsection we will briefly discuss our results for each individual element.\\

\begin{figure}
\resizebox{\hsize}{!}{\includegraphics{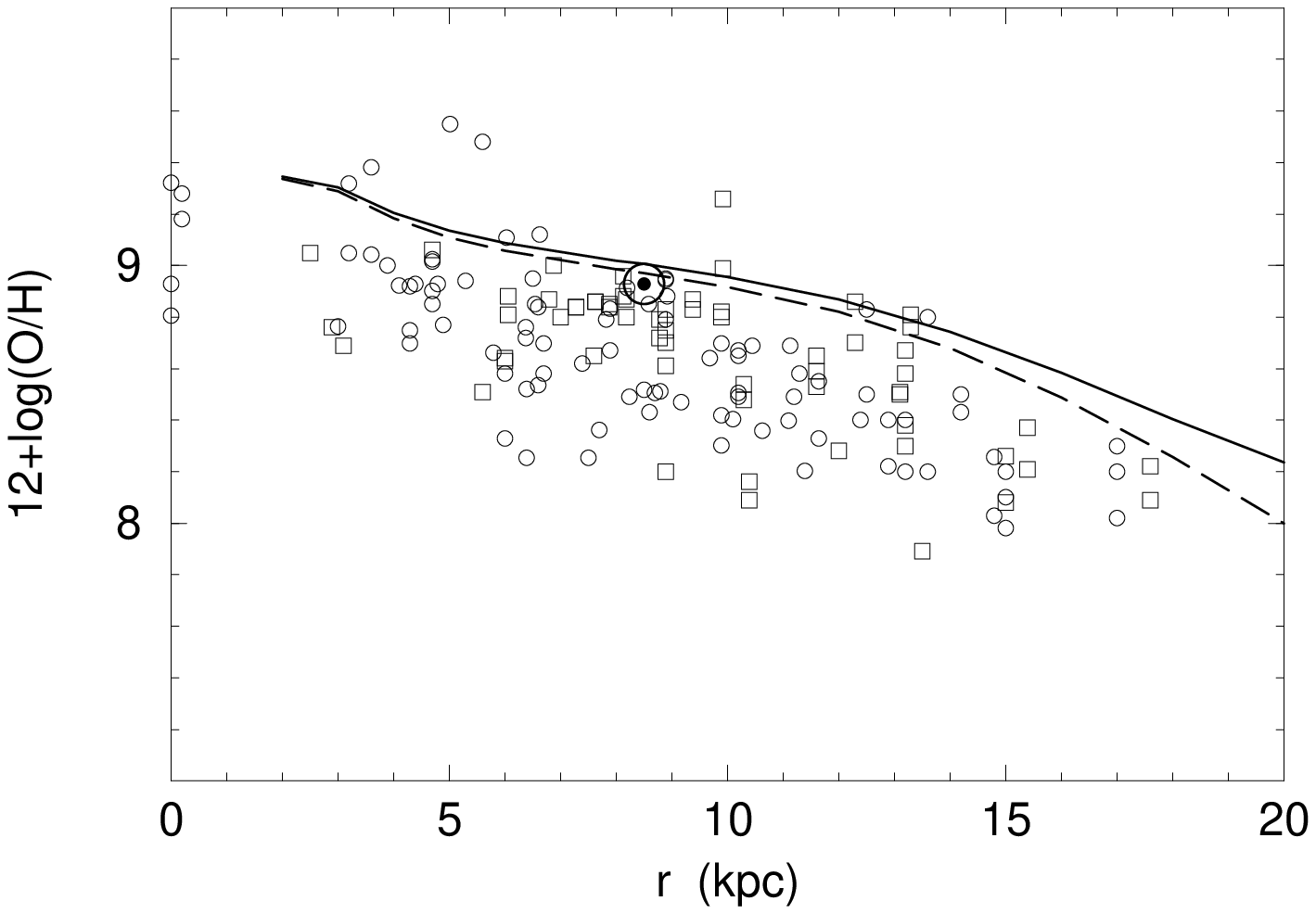}}
\caption{Calculated current O abundance profile in the Milky Way disk. \emph {Solid line}: enriched infall model. \emph {Dashed line}: primordial infall model. Data: \emph {circles}-\ion{H}{ii} regions (Shaver et al. \cite{shav83}; Fich \& Silkey \cite{fich91}; Simpson et al. \cite{simp95}; V\'{\i}lchez \& Esteban \cite{vilc96}; Afflerbach et al. \cite{affl97}; Rudolph \cite{rudo97}; Deharveng et al. \cite{deha00}); \emph {squares}-B stars (Smartt \& Rolleston \cite{smar97}; Gummersbach et al. \cite{gumm98}; Hibbins et al. \cite{hibb98}; Rolleston et al. \cite{roll00}; Smartt et al. \cite{smar01b}). Solar System abundances ($\sun$) are from Anders \& Grevesse (\cite{ande89})}
\label{OH-R}
\end{figure}

\noindent{\bf Oxygen.-} Because O is a good metallicity indicator its radial abundance profile in the Milky Way disk, as well as in external disk galaxies, has been extensively studied. Despite of early discrepancies between results from galactic \ion{H}{ii} regions and PN, which give steep gradients, and observations of B stars, that showed much flatter gradients,  there seems nowadays to be a general convergence towards an abundance gradient with a unique slope $d\log(O/H)/dr\sim-0.07\pm0.01$ dex kpc$^{-1}$ (see Table \ref{hiibpn}). However, there are some hints that the situation might be less clear. The O gradient in the inner Galaxy as traced by young B stars appears to be rather flat (Smartt \cite{smar01a}) in contradiction with results from \ion{H}{ii} regions (Afflerbach et al. \cite{affl97}). The reasons for this discrepancy are unclear, specially so in view that the other elements considered in the same paper do not show such inner flattening. A further conflict comes from the recent work of Deharveng et al. (\cite{deha00}) concerning \ion{H}{ii} regions (an indicator that traditionally has given steep gradients). These authors have collected a new sample of \ion{H}{ii} regions located at galactocentric distances in the range 6.6 to 17.7 kpc, and determined O/H abundances assuming a two--temperature model for the ionized nebula. They obtain a much shallower O gradient (by almost a factor of two) than previous surveys of \ion{H}{ii} regions.

\begin{figure}
\resizebox{\hsize}{!}{\includegraphics{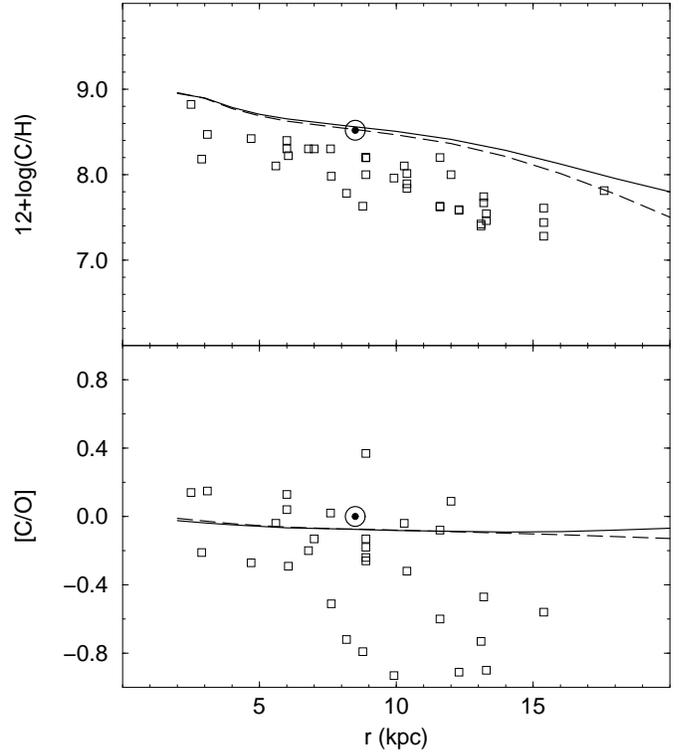}}
\caption{Present C abundance and C abundance ratio to O (referred to the solar value) profiles. \emph {Solid line}: enriched infall model. \emph {Dashed line}: primordial infall model. Data: \emph {squares}-B stars (Gummersbach et al. \cite{gumm98}; Hibbins et al. \cite{hibb98}; Smartt et al. \cite{smar01b} ). Solar System abundances ($\sun$) are from Anders \& Grevesse (\cite{ande89})}
\label{CH-R}
\end{figure}

Our models reasonably reproduce the O abundance profile as displayed in Fig. \ref{OH-R}. There is an small zero--point discrepancy, as stated above, and we get slightly shallower gradients than those derived from B stars. The enriched model gives a result in good agreement with the value quoted in Deharveng et al. (\cite{deha00}). As it is the case for the other elements, the primordial infall model yields a steeper slope, which is still a bit smoother than the canonical value, but compatible with observations, in particular those of young PN. In both models the O profile in the inner disk is flatter than in the external zones.\\

\begin{figure}
\resizebox{\hsize}{!}{\includegraphics{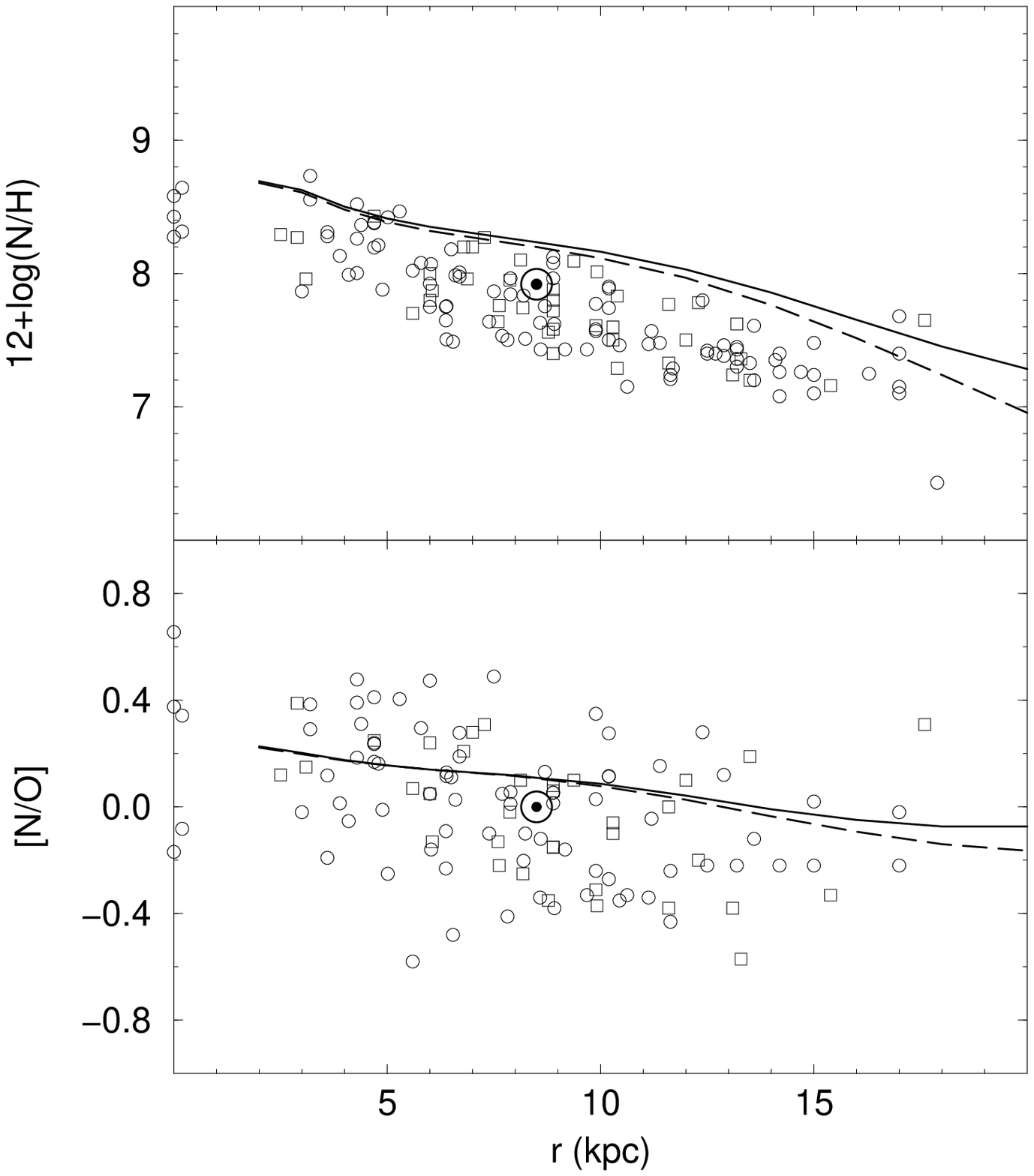}}
\caption{Present N abundance and N/O profiles. \emph {Solid line}: enriched infall model. \emph {Dashed line}: primordial infall model. Data: \emph{circles}-\ion{H}{ii} regions (Shaver et al. \cite{shav83}; Fich \& Silkey \cite{fich91}; Simpson et al. \cite{simp95}; V\'{\i}lchez \& Esteban \cite{vilc96}; Afflerbach et al. \cite{affl97}; Rudolph \cite{rudo97}); \emph {squares}-B stars (Gummersbach et al. \cite{gumm98}; Rolleston et al. \cite{roll00}; Smartt et al. \cite{smar01b}). Solar System abundances ($\sun$) are from Anders \& Grevesse (\cite{ande89})}
\label{NH-R}
\end{figure}

\noindent{\bf Carbon.-} Measures of C abundances in young disk objects are few and rather  uncertain. Observations of \ion{H}{ii} nebulae in a small range of galactocentric distances (6--9 kpc) by Esteban et al. (\cite{este99}) give a steep carbon radial gradient of $-0.133\pm0.022$ dex kpc$^{-1}$. Most of the additional data come from B stars. Gummersbach et al. (\cite{gumm98}) found a much smoother gradient of $-0.045\pm0.014$ dex kpc$^{-1}$. However, later studies by Rolleston et al. (\cite{roll00}) and Smartt et al. (\cite{smar01b}) extending from the inner galaxy to 18 kpc gave gradients of the order of $-0.06\pm0.02$ dex kpc$^{-1}$. Results for the C abundance distribution and the corresponding [C/O] profile are presented in Fig. \ref{CH-R}. Our enriched model produces $d\log(C/H)/dr\sim-0.05$ dex kpc$^{-1}$, while in the calculation with a primordial composition of the accreted matter we obtain a steeper value of $\sim-0.06$ dex kpc$^{-1}$, in excellent accord with the recent data sets. The local carbon abundance at $t_{\sun} = 8.5$ Gyr is nearly solar, which shows that contributions from low-- and intermediate--mass stars are enough to complement the carbon produced by the WW\cite{woos95} yields for massive stars.   
\begin{figure*}
\centering\includegraphics[width=17cm]{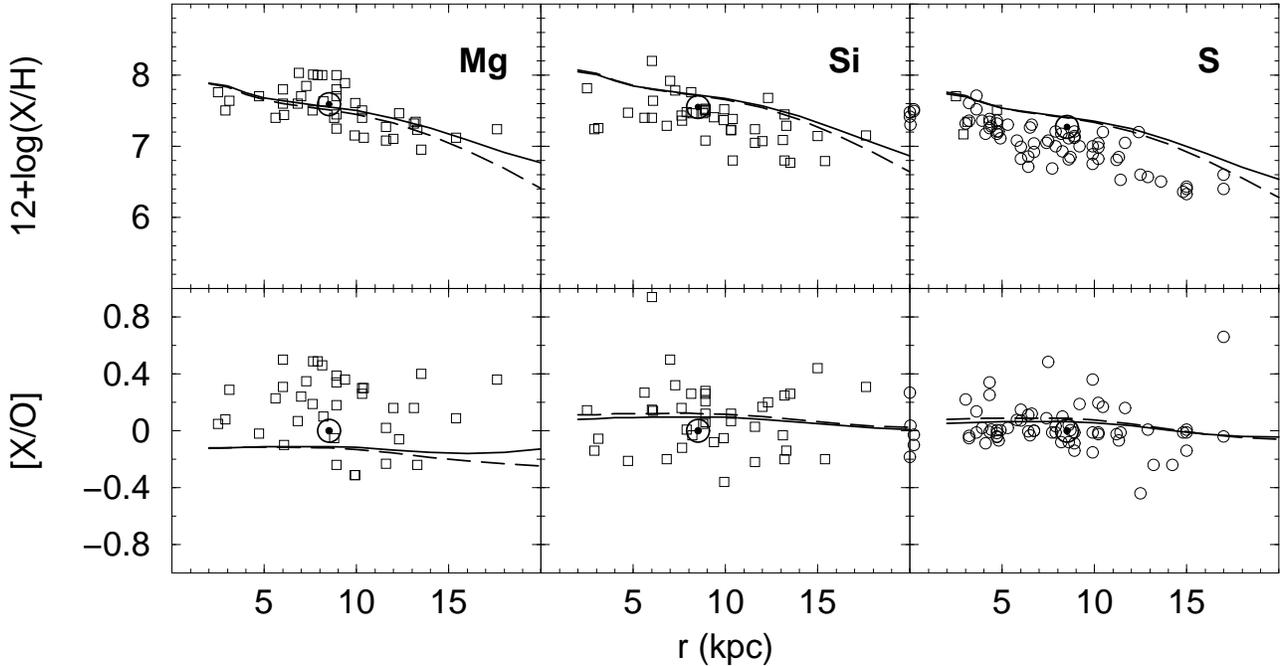}
\caption{Present Mg, Si and S abundance and abundance ratio to O profiles. \emph {Solid line}: enriched infall model. \emph {Dashed line}: primordial infall model. Data: \emph {circles}-\ion{H}{ii} regions (Afflerbach et al. \cite{affl97}); \emph {squares}-B stars (Smartt et al. \cite{smar01b}). Solar System abundances ($\sun$) are from Anders \& Grevesse (\cite{ande89})}
\label{MgSiSH-R}
\end{figure*}

Nevertheless, our models produce a flat C/O profile (see the lower panel of Fig. \ref{CH-R}) because carbon production hardly depends on metallicity (or even decreases as the metallicity grows) in the HG\cite{vand97} yields. Measures of this ratio in local low--mass stars point to an increase of C/O with metallicity (Gustafsson et al. \cite{gust99}). In fact, in the recent study of a sample of B stars by Smartt et al. (\cite{smar01b}) a gradient of $-0.05\pm0.02$ for C/O is found, steeper than the $-0.02\pm0.03$ previously measured by the same group (Rolleston et al. \cite{roll00}), owing to the inclusion of four galactic center stars, carbon--rich but not oxygen--rich, in the sample of Rolleston et al. (\cite{roll00}). In this last paper, the authors themselves notice that their absolute [C/O] ratios are significantly offset from the Orion and solar values. They attribute this discrepancy to errors in LTE C and O abundance determinations that do not invalidate the underlying trend of their [C/O] ratios. If this trend is real, since it cannot be explained by the contribution to carbon abundances by intermediate--mass stars, we must turn to massive stellar yields that include mass loss, such of those of Maeder (\cite{maed92}), where C is produced at a higher rate than O in metal--rich stars due to the enhanced mass--loss in the post He--burning phase. Recent papers by Henry et al. (\cite{henr00}) and by Hou et al. (\cite{hou00}) have shown that metallicity--enhanced carbon yields from massive stars \emph{\`a la} Maeder (\cite{maed92}) could account for the observed carbon abundances in the galactic disk, producing at the same time C/O profiles increasing towards the galactic center. In any case, to settle the question of the role played by stars of different mass ranges in carbon enrichment, more extensive and accurate determinations of the C/O ratio across the galactic disk are needed.\\

\begin{figure*}
\centering\includegraphics[width=17cm]{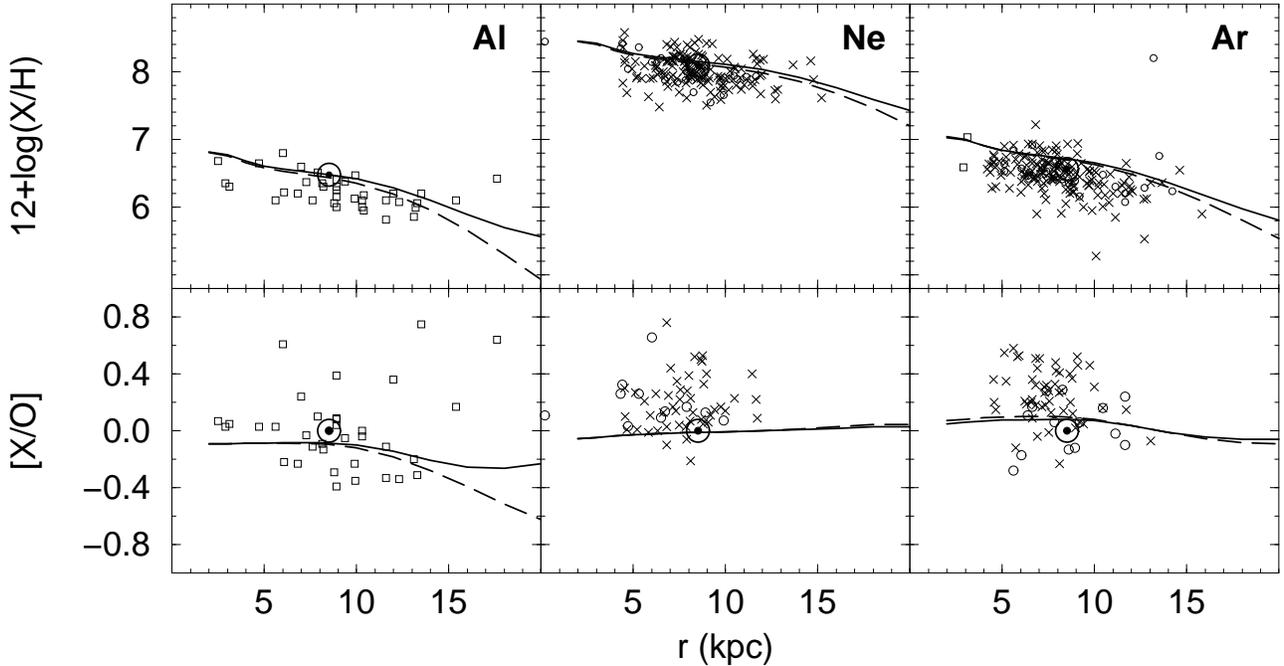}
\caption{Present Al, Mg and Ar abundance and abundance ratio to O profiles. \emph {Solid line}: enriched infall model. \emph {Dashed line}: primordial infall model. Data: \emph {circles}-\ion{H}{ii} regions (Shaver et al. \cite{shav83}; Simpson et al. \cite{simp95}); \emph {squares}-B stars (Gummersbach et al. \cite{gumm98}; Rolleston et al. \cite{roll00}; Smartt et al. \cite{smar01b}). Solar System abundances ($\sun$) are from Anders \& Grevesse (\cite{ande89})}
\label{AlNeArH-R}
\end{figure*}

\noindent{\bf Nitrogen.-} N has been extensively observed in \ion{H}{ii} regions, B stars and PNII (see Table \ref{hiibpn}). In general, its gradient seems steeper ($\sim-0.08$ dex kpc$^{-1}$) than that of O, but there is a wide scatter in the values quoted in the literature, going from $-0.05$ dex kpc$^{-1}$, as determined by Pasquali \& Perinotto (\cite{pasq93}) in PN, to $-0.11$ dex kpc$^{-1}$ in the study of \ion{H}{ii} regions by Rudolph et al. (\cite{rudo97}). The recent analysis by Smartt et al. (\cite{smar01b}), including N abundance determinations in four young B--type stars lying towards the Galactic Center in the sample of Rolleston et al. (\cite{roll00}), derived a N gradient of $-0.06\pm0.02$ dex kpc$^{-1}$. In our calculations we also obtain a steeper N gradient as compared with the O profile, with values $\sim-0.07$ dex kpc$^{-1}$ (as always, slightly steeper in the primordial infall model), in good agreement with the observational determinations. Low-- and intermediate--mass stars contributions are mandatory to match the local abundance, as pointed out by Hou et al. (\cite{hou00}) who, including only yields from massive stars, found subsolar N abundance in the solar ring, even when the evolution was calculated with the Maeder (\cite{maed92}) stellar yields.

The ratio N/O is essential to understand the relative importance of primary versus secondary production of N. A clear tendency of N/O ratios increasing with metallicity (i.e. with decreasing galactocentric radius) would indicate that secondary production dominates de N evolution. Observations of external spiral galaxies (Vila-Costas \& Edmunds \cite{vila93}; Henry et al. \cite{henr00}) show that at metallicities higher than $12+\log(O/H)\sim 8.3$ nitrogen behaves as secondary. In the Galaxy, \ion{H}{ii} regions data clearly show higher than solar N/O ratios in the inner Galaxy, but nearly solar values in the outer disk (V\'{\i}lchez \& Esteban \cite{vilc96}; Rudolph et al. \cite{rudo97}). However, observations of B stars showed a unique and rather steep slope, from $-0.04\pm0.02$ dex kpc$^{-1}$ (Rolleston et al. \cite{roll00}) to $-0.06$ dex kpc$^{-1}$ (Smartt et al. \cite{smar01b}), for the N/O radial profile. Contribution to nitrogen abundances by low-- and intermediate--mass stars allows, in our models, to rightly fit the solar value. The N yields of HG\cite{vand97} present a moderate secondary behaviour, that translates into a more pronounced N gradient than in the case of O. Therefore, our models give N/O profiles increasing towards the inner Galaxy, characterized by a gradient in N/O of $\sim-0.02$ dex kpc$^{-1}$, as shown in the lower panel of Fig. \ref{NH-R}. Nevertheless, this result is still much less than the values derived from B stars. Thus, our calculations support the idea that intermediate--mass stars give an important contribution to the building up of the nitrogen abundance in the Galaxy, but new measurements of the N/O ratio are crucial to asses whether the nitrogen synthesis in those stars has a clear secondary character. \\

\noindent{\bf The $\alpha$--elements Magnesium, Silicon and Sulphur.-} These three $\alpha$--elements should correlate with O since all of them are produced and returned to the ISM through Type II supernova explosions. The observed radial profiles of Mg and Si have been obtained from B stars surveys, and they are well fitted by linear relationships with slopes of $\sim-0.07\pm0.01$ and $\sim-0.06\pm0.01$ dex kpc$^{-1}$, respectively, compatible with the O gradient at 1 $\sigma$ level. S has been observed in \ion{H}{ii} regions and PN, and its gradient is also similar to that of O. Our results show a nice agreement with the data (see upper panels of Fig. \ref{MgSiSH-R} and Table \ref{hiibpn}), in particular in the case of our primordial infall model, which again gives steeper gradients than the enriched model. 

We display our results for the [X/O] ratios for the three elements in the lower panels of Fig. \ref{MgSiSH-R}. As expected on theoretical grounds, their calculated abundance ratios to O along the galactic disk are flat, with solar absolute values for Si and S, while in the case of the Mg/O ratio we obtain slightly subsolar values in the whole disk, although the difference is not large. As pointed out in ALC\cite{alib01}, and also by Goswami \& Prantzos (\cite{gosw00}), the WW\cite{woos95} yields underproduce Mg, showing, besides, unexplained metallicity effects.\\  

\noindent{\bf Aluminum, Neon and Argon.-} Our results for these elements are presented in Fig. \ref{AlNeArH-R} (upper panels: gradients; lower panels: [X/O] ratios). Aluminum has been exclusively measured in B--type stars. Gummersbach et al. (\cite{gumm98}), Rolleston et al. (\cite{roll00}) and Smartt et al. (\cite{smar01b}) found relatively shallow Al gradients, of the order of $-0.05\pm0.01$ dex kpc$^{-1}$. This is rather surprising, since Al is an odd monoatomic element (\element[][27]{Al}), subjected to the well known odd--even effect, i.e. stellar yields increasing with metallicity. Therefore, owing to the low metal enrichment of the outer galaxy, the Al abundances there should increase more slowly than those of pure primary elements like O, giving rise to a rather steep gradient. In fact, that is precisely the behaviour that we obtain in our models (see Fig. \ref{AlNeArH-R}). Owing to the marked odd--even effect in the Al yields of WW\cite{woos95}, we produce an Al gradient steeper than those of O and the $\alpha$--elements. As a consequence,  although our calculated present local Al/O ratio is almost solar, it shows a noticeable dependence on galactocentric distance, and hence metallicity, in the outer Galaxy (more evident in the case of our primordial infall model), again at odds with the data that show no trend at all. Similar results, with even a more pronounced Al gradient, were obtained by Hou et al. (\cite{hou00}), who also employed the WW\cite{woos95} yields for massive stars. In view of the above, one is forced to conclude that theory and observations do not match each other, and that the current Al yields for massive stars should be revised. 

Ne and Ar have been mainly measured in PN, and in much lesser extent in \ion{H}{ii} regions. The large set of data from PN (Maciel \& K\"oppen \cite{maci94}; Maciel \& Quireza \cite{maci99}) give a Ne gradient of $\sim-0.04$ dex kpc$^{-1}$, and a somewhat larger value of $\sim-0.05$ dex kpc$^{-1}$ for the Ar gradient. Our models produce results for these two elements in very good agreement with the observations, and comparable to those calculated for the rest of the $\alpha$--elements. The corresponding ratios to O are $\sim$ solar in the local ring, with the expected flat distribution. Nevertheless, the Ne/O profile remains below most of the observations. Regarding this last point, we remind that  PNI are probably O--poor, and that measurements of Ne/O in PNI could be overestimated (Peimbert et al. \cite{peim95}).\\   

\begin{figure}
\resizebox{\hsize}{!}{\includegraphics{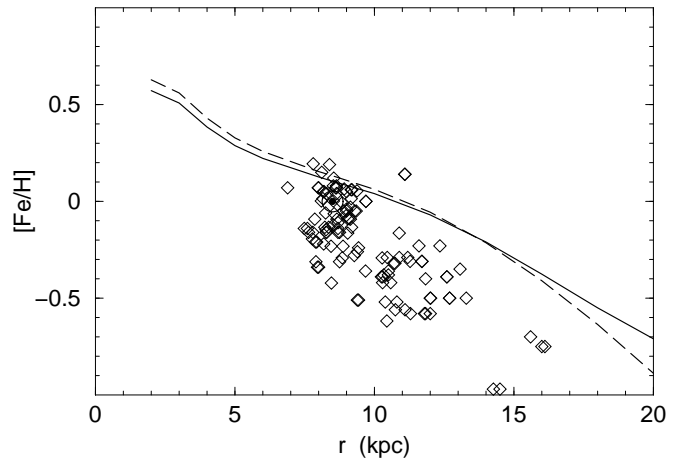}}
\caption{Calculated iron profile compared with observations in open clusters. \emph {Solid line}: enriched infall model. \emph {Dashed line}: primordial infall model. Data: Cameron (\cite{came85}), Friel (\cite{frie95}) and Carraro et el. (\cite{carr98})}
\label{FeH-R}
\end{figure}

\noindent{\bf Iron.-} In Fig. \ref{FeH-R} we present the calculated Fe profile compared to the current data, while the values of  the corresponding slopes appear in Table \ref{oclu}. Fe abundances are usually measured in open clusters. Earlier determinations of its radial gradient gave values in the range $-0.06$ to $-0.09$ dex kpc$^{-1}$. Although there have been claims that a two--step distribution with an abrupt discontinuity at a galactocentric distance of $\sim10$ kpc would best adjust the observations (Twarog et al. \cite{twar97}), recent work by Carraro et al. (\cite{carr98}) and Friel (\cite{frie99}) support linear radial Fe profiles, with a slope similar to that found in the former studies. We obtain a Fe gradient of $\sim-0.06$ dex kpc$^{-1}$, in excellent agreement with the value quoted by Friel (\cite{frie99}). It is disputable to what extent open clusters trace the current abundances profiles, since those objects span a rather wide interval of ages. Nevertheless, when studying open clusters samples of different ages there is a general agreement on no or just very little evolution of the Fe gradient during the galactic disk lifetime. Actually, however, the absence of any age--metallicity relation for the galactic open clusters, in clear contradiction with the local field stars, casts some doubts on our understanding of the Fe evolution in the Galaxy.

\subsection{Gradient evolution \label{gradevol}}

A crucial point relative to the gradients of elemental abundances in the Galaxy, that can help to discriminate between theoretical models, is their time evolution. Have they been steepening or flattening? Or we just have had a constant gradient during most of the Galaxy evolution? As indicated in the Introduction, most models that successfully account for the majority of the chemical features of the solar neighborhood, as well as  of the present galactic disk, find a gradual flattening with time of the radial abundance distributions (e.g. Portinari \& Chiosi \cite{port99}; Hou et al. \cite{hou00}). Nonetheless, there also exist models that give abundance gradients that progressively steepen with time. For instance, Chiappini et al. (\cite{chia01}), with a model involving star formation thresholds for both halo and disk, obtain abundance gradients that increase in the course of galactic evolution. Among other reasons, this opposition between predictions from different models comes basically from the diversity in the prescriptions used for the time dependence of the SFR and the infall. 

Unfortunately, the present status of observations can not help very much in solving the controversy. We have already mentioned that abundance measurements obtained from PN or open clusters could give some insights on this subject, owing to the relatively wide range of ages covered by those objects. Maciel \& Quireza (\cite{maci99}) found that abundance gradients derived from PN are shallower by 0.01--0.03 dex kpc$^{-1}$ than those obtained from young objects, and roughly estimated a slowly steepening rate of $-0.004$ dex kpc$^{-1}$ Gyr$^{-1}$ for the oxygen gradient, but large uncertainties are involved in this result. First, the ages, masses and distances of PN are crudely determined. Second, abundances in PN might not be truly representative of the composition of the ISM from which their parent stars formed. Finally, the distribution of PN across the disk could have been altered by orbital diffusion. 

The same way as the age--metallicity relation for nearby stars is a record of the metal enrichment in the local Galaxy, Fe measures in open clusters with a variety of ages do reveal some of the past history of the galactic abundance profiles. Besides, it is expected that orbital diffusion hardly affects the open cluster population, and their ages are more reliably determined than those of PN. Most studies indicate that the mean iron cluster abundance shows no dependence on age, pointing to a prompt initial enrichment of the disk (Friel \cite{frie99}). This trend, i.e. that the radial iron gradient has not or just slightly changed with time, is found as well in Carraro et al. (\cite{carr98}). In fact, this last work hints to somewhat steeper gradients in the past, for intermediate--age clusters, although due to the small sample analyzed this result has little statistical significance. Results in the same direction have been preliminarily obtained by Bragaglia et al. (\cite{brag99}), though again with poor statistics. In conclusion, open clusters temptatively indicate that the radial gradients have remained constant, or even slightly flattened, during most of the disk lifetime.

\begin{figure}
\resizebox{\hsize}{!}{\includegraphics{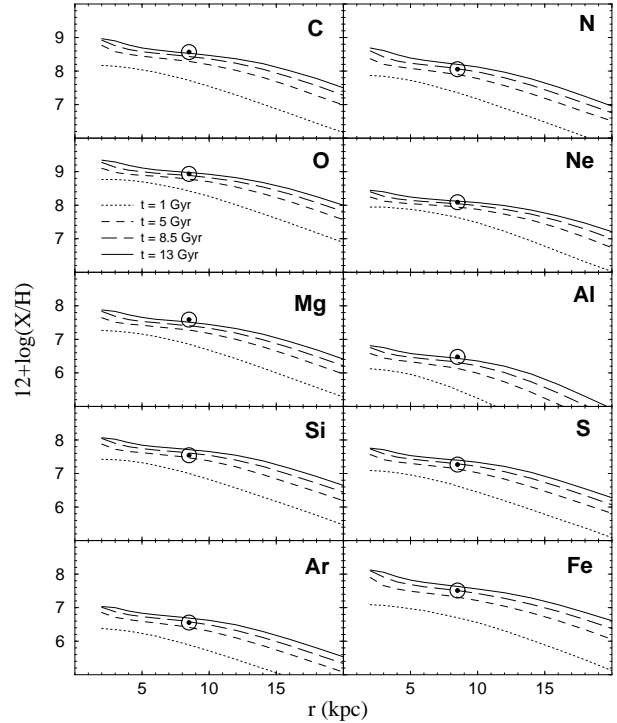}}
%{\par\centering \resizebox*{7.5cm}{!}{\includegraphics{grf/abund5x2.DRdp.eps}} \par}
\caption{Abundance profiles for different times (1, 5, 8.5 and 13 Gyr) in our primordial infall model. Long dashed lines correspond to the time of formation of the Sun. Solar system abundances ($\sun$) from Anders \& Grevesse (\cite{ande89})}
\label{evolabund}
\end{figure}

We display in Fig. \ref{evolabund} the radial distribution of the ten elements that we are discussing, obtained with the primordial infall model, at different times. We do not present our results for the enriched model since they are quite similar, apart from giving, as mentioned above, shallower gradients in the outer Galaxy. The gradient evolution is more clearly seen in Fig. \ref{evolgrad}, where the explicit time evolution of the abundance gradients in the 4--16 kpc region is shown, both for the enriched (\emph{solid line}) and the primordial (\emph{dashed line}) models. We see a rapid flattening of the gradients for the majority of the elements, with the main changes in the radial gradients being produced during the first two Gyrs. Obviously, this trend is most noticeably when we consider enriched infall, since at early times the metallicity of the matter that settles into the galactic disk quickly erases the initial gradients. The evolution is almost completed at 5 Gyr. Afterwards, the flattening proceeds at a very slow pace till reaching the current values listed in Tables \ref{hiibpn} and \ref{oclu}. We must keep in mind that the enriched infall is a kind of upper limit for the metallicity of the accreted material. A value of $0.1\, Z_{\sun}$ is what is observed nowadays, but the infall could have been metal--poorer in the past. In such a case, the gradient evolution would be something in--between those displayed for the two extreme compositions considered. The abundance profiles in the inner Galaxy remain flatter during most of the evolution, due to the initially high infall and star formation rates, which induce a prompt saturation of the ISM, whose metallicity practically does not increase in later epochs (besides the fact that the stellar yields that we use do stop at solar metallicity, as already mentioned). 

\begin{figure}
\resizebox{\hsize}{!}{\includegraphics{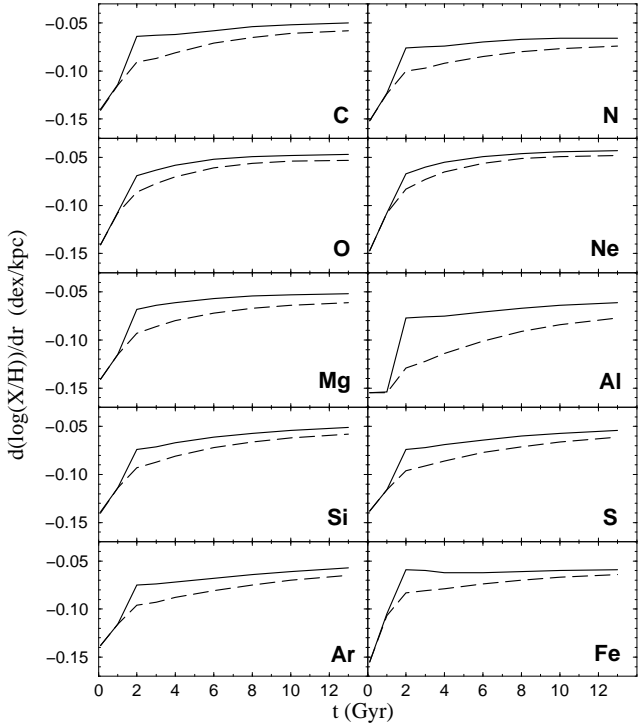}}
\caption{Evolution of the gradients in the range 4--16 kpc. \emph{Solid line}: enriched infall model. \emph{Dashed line}: primordial infall model. A quick flattening of the gradient is obtained in the enriched infall model before 2 Gyr, due to the accretion of $0.1\, Z_{\sun}$ material. At $t_{\mathrm{G}}$ it still produces smoother gradients than the primordial infall model}
\label{evolgrad}
\end{figure}
 
The aluminum gradient evolution in the primordial infall model is clearly different from the rest, in the sense that the gradient steadily decreases during the whole evolution. The reason is once again the aluminum nature by being an odd--even element, with yields that increase with metallicity. This effect is seen as a slower evolution of the gradient as a consequence of the idle enrichment in the outer Galaxy.

\begin{figure}
\resizebox{\hsize}{!}{\includegraphics{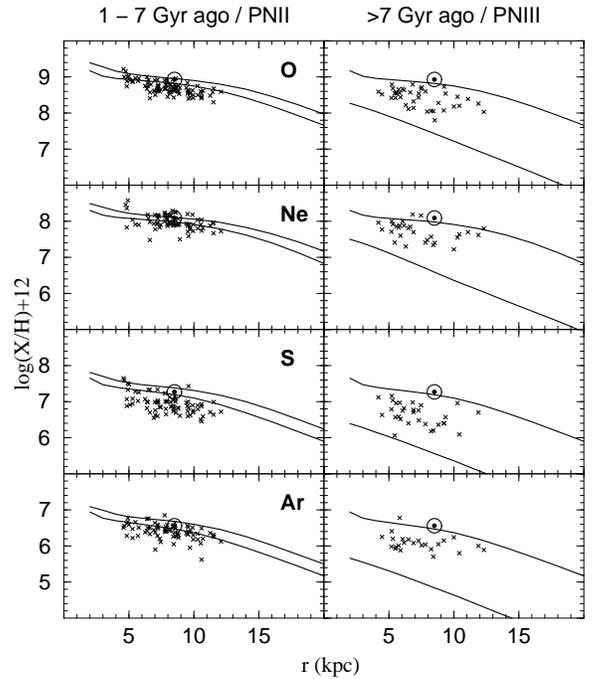}}
\caption{Abundance profiles obtained in our primordial infall model at different times compared with data from PN. Left panels correspond to PNII data (with ages between 1 and 7 Gyr) and the abundance profiles are from our models at $t = 6$ and 12 Gyr. Right panels are for PNIII observations (ages $\geq 7$ Gyr) and model results for $t = 0.1$ and 6 Gyr. PN abundances from Maciel \& K\"oppen (\cite{maci94}) and Solar System abundances ($\sun$) from Anders \& Grevesse (\cite{ande89})}
\label{PN}
\end{figure}

In summary, our models, as it is the case in some previous work on this topic, produce abundance gradients that tend to decrease in the course of the Galaxy lifetime, basically in its earlier stages. This result is in concordance with the indications about the Fe abundance behaviour as deduced from studies of open clusters, but it is in contradiction with evidences obtained from PN. Actually, the current observational status on the temporal evolution of the chemical abundances, both in the Galaxy and in other spirals, is subjected to such uncertainties that it is far from allowing us to discriminate whether the galactic abundance profiles steepen or flatten with time. For instance, our model predictions can reasonably accommodate the data from PN. Following the work of Hou et al. (\cite{hou00}), in Fig. \ref{PN} we compare, in the case of the primordial infall model, the abundance profiles at various epochs with the observed distribution in PNII (ages between 1 and 7 Gyr) and PNIII (ages $\geq$ 7 Gyr), for the four elements (O, Ne, S and Ar) that have been measured there. We reach a mild agreement with the data from PNII for Ne and Ar and in less extent for O, while S appears clearly above the data. In any case, the spread of PNII abundances, that could in part  be attributed to the errors associated with distance and age determinations, is a little bit larger than the range of our predictions. The PNIII abundances, however, are well reproduced by our model, as it is apparent in the right panels of Fig. \ref{PN}, where almost all the data points fall between our predicted abundance profiles at times of 6 Gyr (age of 7 Gyr) and 0.1 Gyr (age of 12.9 Gyr).

\section{Summary and conclusions \label{conclusions}}

The chemical model developed in ALC\cite{alib01}, that nicely accounts for the observational characteristics in the solar neighborhood, has been extended to the whole galactic disk. Our model assumes an inside--out scenario for the assembling of the Milky Way disk through infall of external gas, with a timescale that is a linear function of the galactocentric distance. Two different compositions of the accreted material have been considered. In our primordial infall model we assume that the matter that settles down to build the disk has a composition corresponding to the Big Bang nucleosynthesis. We have also investigated models, our enriched infall models, in which the infalling material that forms the thin disk has been preenriched up to a metallicity of $0.1\,Z_{\sun}$, in line with metallicity measurements of High Velocity Clouds. The adopted star formation rate has an explicit dependence on the total surface mass density, so that it implicitly becomes a function of the radial distance. Using metallicity--dependent yields for the whole interval of stellar masses, we have calculated the evolution of all the stable isotopes between hydrogen and zinc, although in the present paper we only analyze the elements that have been observed in the Galaxy, for which we have collected from the literature a wide and recent set of data. We have concentrated on the study of the radially--dependent properties of the galactic disk, giving special emphasis to the abundance profiles. Our models do not include radial flows because we prefer to keep as short as possible the number of adjustable parameters (but see Portinari \& Chiosi \cite{port00} for a recent discussion on the effects of radial flows in standard chemical models).

We briefly resume our principal conclusions as follows:

\begin{itemize}
\item We obtain satisfactory fits to the observed gas, star and SFR distributions in the Milky Way disk, although we do not reproduce adequately the molecular ring at 4 kpc, a common feature of models that do not take into account radial flows and/or the effects of a central bar. Nevertheless, we fairly adjust the abundances of gas and the level of star formation in the galactic disk beyond the molecular ring, precisely the region where most of the elemental abundances have been measured and, consequently, the crucial zone for determining the chemical gradients.

\item All the studied elements show noticeable gradients at the present epoch, whose values are compatible with the data from young objects, like \ion{H}{ii} regions and B--type stars, at 1$\sigma$ level. For the best observed element, oxygen, our calculated gradient is a little bit flatter than the measured one, but still within the observational limits. Models that include enriched infall give shallower gradients than models that assume a primordial composition, typically by 0.006-0.007 dex kpc$^{-1}$, since the metal content of the accreted matter levels off the chemical abundances in the outer Galaxy, where the efficiency of star formation remains low along the complete evolution. If the shallow oxygen gradient determined by Deharveng et al. (\cite{deha00}) were confirmed, enriched instead of primordial infall could be a better assumption in theoretical models of the Galaxy evolution. We obtain sharper gradients for N (moderately secondary in the yields of intermediate--mass stars from HG\cite{vand97}) and also for Al (an odd--even element) than for primary elements like O and other $\alpha$--elements.

\item The primary elements have current profiles of their ratios to oxygen that show no trend with galactocentric radius (and thus, with metallicity), in agreement with the available data and with theoretical expectations. 

\item Contributions from intermediate--mass stars to C and N are enough to reproduce the observed absolute abundances of these elements. However, we get a flat C/O ratio, which seems to be in contradiction with recent observations of \ion{H}{ii} regions (Henry \& Worthey \cite{henr99}) and B stars (Smartt et al. \cite{smar01b}), which indicate an increase of C/O with metal contents. Such trend can be reproduced in calculations that include the effects of stellar winds on the yields of massive stars, as shown by Hou et al. (\cite{hou00}), but in that case the absolute C abundances would be overestimated when adding the contribution from intermediate--mass stars. Owing to the mild secondary character of N in the yields from HG\cite{vand97}, our models produce a definite gradient in the N/O distribution, although lower than indicated by observations.

\item Our model, as it also happens in other work that adopts similar assumptions, predicts that abundance gradients in the disk do flatten in the course of galactic evolution. If the material that settles into the disk has an enriched composition, most of this flattening occurs during the firsts 2 Gyr, while this period is more extended when we consider primordial infall. Later on, the gradients continue to flatten but in a quite gentle manner. In other words, they exhibit little evolution in recent times. Secondary elements depart to some extent from this behaviour, in the sense that their gradients do not hastily smooth out, and they show a more steady pace towards their present values.   

\end{itemize}


\begin{thebibliography}{}
\bibitem[1997]{affl97}Afflerbach, A., Churchwell, E., \& Werner, M.W. 1997, ApJ, 478, 190
\bibitem[1998]{allen98}Allen, C., Carigi, \& L., Peimbert M. 1998, ApJ, 494, 247
\bibitem[2001]{alib01}Alib\'es, A., Labay, J., \& Canal, R. 2001, A\&A, 370, 1103 (ALC2001)
\bibitem[1989]{ande89}Anders, E., \& Grevesse, N. 1989, Geochim. Cosmochim. Acta, 53, 197
\bibitem[1999]{berc99}Berczik, P. 1999, A\&A, 348, 371
\bibitem[1999]{bois99}Boissier, S., \& Prantzos, N. 1999, MNRAS, 307, 857
\bibitem[1999]{brag99}Bragaglia, A., Tosi, M., Marconi, G., \& Carretta E. 1999, astro-ph/9912130
\bibitem[1985]{came85}Cameron, L.M. 1985, A\&A, 147, 47
\bibitem[1998]{carr98}Carraro, G., Ng, Y.K., \& Portinari, L. 1998, MNRAS, 296, 1045
\bibitem[1999]{chan99}Chang, R.X., Hou, J.L., Shu, C.G., \& Fu, C.Q. 1999, A\&A, 350, 38
\bibitem[1997]{chia97}Chiappini, C., Matteucci, F., \& Gratton, G. 1997, ApJ, 477, 765
\bibitem[2000]{chia00}Chiappini, C., Matteucci, F., \& Padoan, P. 2000, ApJ, 528, 711
\bibitem[2001]{chia01}Chiappini, C., Matteucci, F., \& Romano, D. 2001, astro-ph/0102134
\bibitem[1993]{dame93}Dame, T.M. 1993, in Back to the Galaxy, (Holt S., Verter F., eds.), 267
\bibitem[2000]{deha00}Deharveng, L., Pe\~na, M., Caplan, J., \& Costero R. 2000, MNRAS, 311, 329
\bibitem[1994]{dopi94}Dopita, M.A., \& Ryder, S.D. 1994, ApJ, 430, 163
\bibitem[1993]{edva93}Edvardsson, B., Anderson, J., Gustafsson, B., Lambert, D.L., Nissen, P.E., \& Tomkin, J. 1993, A\&A, 275, 101
\bibitem[2001]{eise01}Eisenhauer, F. 2001, in Starbursts: Near and Far (eds. 
Tarconi, L.J. Lutz, D.), (in press) astro-ph/0101321
\bibitem[1999]{este99}Esteban, C., Peimbert, M., Torres-Peimbert, S., \& Garc\'{\i}a-Rojas, J. 1999, Rev. Mex. Astron. Astrof., 35, 65
\bibitem[1991]{fich91}Fich, M., \& Silkey, M. 1991, ApJ, 366, 107
\bibitem[1992]{fitz92}Fitzsimmons, A., Dufton, P.L., \& Rolleston, W.R.J. 1992, MNRAS, 259, 489
\bibitem[1997]{fore97}Forestini, M., \& Charbonnel, C. 1997, A\&AS, 123, 241
\bibitem[1998]{freu98}Freudenreich, H. 1998, ApJ, 492, 495
\bibitem[1993]{frie93}Friel, E.D., \& Janes, K.A. 1993, A\&A, 267, 75 
\bibitem[1995]{frie95}Friel, E.D. 1995, ARA\&A, 33, 381
\bibitem[1999]{frie99}Friel, E.D. 1999, Ap\&SS, 265, 271
\bibitem[1985]{gehr85}Gehren, T., Nissen, P.E., Kudritzki, R.P., \& Butler, K. 1985, in Production and Distribution of CNO Elements, ed. I.J. Danzinger, F. Matteucci, \& K. Kjaer (Garching: ESO), 171
\bibitem[1989]{gilm89}Gilmore, G., Wyse, R., \& Kuijen, K. 1989, in Evolutionary Phenomena in Galaxies, J. Beckman, \& B. Pagel (eds.), Cambridge University Press, 172
\bibitem[2000]{gosw00}Goswami, A., \& Prantzos, N. 2000, A\&A, 359, 191
\bibitem[1978]{gui78}Guibert, J., Lequeux, J., \& Viallefond, F. 1987, A\&A, 68, 1
\bibitem[1998]{gumm98}Gummersbach, C.A., Kaufer, A., Sch\"afer, D.R., Szeifert, T., \& Wolf, B. 1998, A\&A, 338, 881
\bibitem[1999]{gust99}Gustafsson B., Karlsson T., Olsson E., Edvarsson B. \& Ryde N. 1999, A\&A, 342, 426 
\bibitem[1982]{gust82}G\"usten, R., \& Mezger, M. 1982, Vistas in Astr., 26, 159
\bibitem[1999]{henr99}Henry, R.B.C., \& Worthey, G. 1999, PASP, 111, 919 
\bibitem[2000]{henr00}Henry, R.B.C., Edmunds, M.G., \& K\"oppen, J. 2000, ApJ, 541, 660 
\bibitem[1998]{hibb98}Hibbins, R.E., Dufton, P.L., Smartt, S.J., \& Rolleston, W.R.J. 1998, A\&A, 332, 681
\bibitem[2000]{hou00}Hou, J.L., Prantzos, N., \& Boissier, S. 2000, A\&A, 362, 921
\bibitem[1979]{jane79}Janes, K.A. 1979, ApJS, 39, 135
\bibitem[1998]{jose98}Jos\'e, J., \& Hernanz, M. 1998, ApJ, 494, 680
\bibitem[1994]{kauf94}Kaufer, A., Szeifert, T., Krenzin, R., Baschek, B., \& Wolf, B. 1994, A\&A, 289, 740
\bibitem[1998]{kenn98}Kennicutt, R. 1998, ApJ, 498, 541
\bibitem[1994]{kili94}Kilian-Montenbruck, J., Gehren, T., \& Nissen, P.E. 1994, A\&A, 291, 757
\bibitem[1993]{krou93}Kroupa, P., Tout, C., \& Gilmore, G. 1993, MNRAS, 262, 545
\bibitem[1976]{Larson76}Larson, R.B. 1976, MNRAS, 176, 31
\bibitem[2000]{limo00}Limongi, M., Straniero, O., \& Chieffi, A. 2000, ApJS, 129, 625
 \bibitem[1985]{lyne85}Lyne, A.G., Manchester, R.N., \& Taylor, J.H. 1985, MNRAS, 213, 613
\bibitem[1994]{maci94}Maciel, W.J., \& K\"oppen, J. 1994, A\&A, 282, 436
\bibitem[1999]{maci99}Maciel, W.J., \& Quireza, C. 1999, A\&A, 345, 629
\bibitem[1987]{maed87}Maeder, A. 1987, A\&A, 178, 159
\bibitem[1992]{maed92}Maeder, A. 1992, A\&A, 264, 105
\bibitem[1996]{mari96}Marigo, P., Bressan, A., \& Chiosi, C. 1996, A\&A, 313, 545
\bibitem[2001]{mari01}Marigo, P. 2001, A\&A, 371, 152
\bibitem[2000]{mart00}Martins, L.P., \& Viegas, S.M.M. 2000, A\&A, 361, 1121
\bibitem[1997]{molla97}Moll\'a, M., Ferrini, F., \& Diaz, A.I. 1997, ApJ, 475, 519
\bibitem[1993]{pasq93}Pasquali, A., \& Perinotto, M. 1993, A\&A, 280, 581
\bibitem[1978]{peim78}Peimbert, M. 1978, IAU Symp. 76, \emph{Planetary Nebulae}, 215
\bibitem[1995]{peim95}Peimbert, M., Luridana, V., \& Torres-Peimbert S., 1995, Rev. Mex. Astron. \& Astrofis., 31, 147
\bibitem[1995]{piat95}Piatti, A.E., Claria, J.J., \& Abadi, M.G. 1995, AJ, 110, 2813
\bibitem[1999]{port99}Portinari, L., \& Chiosi, C. 1999, A\&A, 350, 827
\bibitem[2000]{port00}Portinari, L., \& Chiosi, C. 2000, A\&A, 355, 929
\bibitem[1991]{rana91}Rana, N.C. 1991, ARA\&A, 29, 12
\bibitem[2000]{roch00}Rocha-Pinto, H.J., Maciel, W.J., Scalo, J., \& Flynn, C. 2000, A\&A, 358, 850
\bibitem[2000]{roll00}Rolleston, W.R.J., Smartt, S.J., Dufton, P.L., \& Ryans, R.S.I. 2000, A\&A, 363, 537
\bibitem[1997]{rudo97}Rudolph, A.L., Simpson, J.P., Haas, M.R., Erickson, E.F., \& Fich, M. 1997, ApJ, 489, 94
\bibitem[1997]{sack97}Sackett, P. 1997, ApJ, 483, 103
\bibitem[1955]{salp55}Salpeter, E. 1955, ApJ, 121, 161
\bibitem[1997]{saml97}Samland, M., Hensler, G., \& Theis, Ch. 1997, ApJ, 476, 544
\bibitem[1986]{scal86}Scalo, J.M. 1986, FCPhys, 11, 1
\bibitem[1959]{schm59}Schmidt, M. 1959, ApJ, 129, 243
\bibitem[1983]{shav83}Shaver, P.A., McGee, R.X., Newton, L.M., Danks, A.C., \& Pottasch, S.R. 1983, MNRAS, 204, 53
\bibitem[1995]{simp95}Simpson, J.P., Colgan, S.W.J., Rubin, R.H., Erickson, E.F., \& Haas, M.R. 1995, ApJ, 444, 721
\bibitem[1997]{smar97}Smartt, S.J., \& Rolleston, W.R.J. 1997, ApJ, 481, L47
\bibitem[2001]{smar01a}Smartt, S.J. 2000, In: Giovannelli, F., Matteucci, F. (eds.) The chemical evolution of the Milky Way: stars versus clusters. Kluwer, vol. 255
\bibitem[2001]{smar01b}Smartt, S.J., Venn, K.A., Dufton, P.L., Lennon, D.J.,  Rolleston, W.R.J., \& Keenan, F.P. 2001, A\&A, 367, 86
\bibitem[1994]{stei94}Steinmetz, M., \& M\"uller, E. 1994, A\&A, 281, L97
\bibitem[1993]{thie93}Thielemann, F.-K., Nomoto, K., \& Hashimoto, M. 1993, in Origin and Evolution of the Elements, ed. N. Prantzos, E. Vangioni-Flam, \& M. Cass\'e (Cambridge: Cambridge Univ. Press), 297
\bibitem[1980]{tins80}Tinsley, B.M. 1980, FCPh, 5, 287
\bibitem[1988]{tosi88}Tosi, M. 1988, A\&A, 197, 47
\bibitem[1997]{twar97}Twarog, B.A., Ashman, K.M., \& Anthony-Twarog, B.J. 1997, AJ, 114, 2556
\bibitem[1997]{vand97}van den Hoek, L.B., \& Groenewegen, M.A.T. 1997, A\&AS, 123, 305 (HG1997)
\bibitem[1993]{vila93}Vila-Costas, M.B., \& Edmunds, M.G. 1993, MNRAS, 265, 199
\bibitem[1996]{vilc96}V\'{\i}lchez, J.M., \& Esteban, C. 1996, MNRAS, 280, 720
\bibitem[1999]{wakk99}Wakker, B.P., Howk, J.C., Savage, B.D., van Woerden, H., Tufte,  S.L., Schwarz, U.J., Benjamin, R., Reynolds, R.J., Peletier, R.F., \& Kalberla, P.M.W. 1999, Nat, 402, 388
\bibitem[1995]{woos95}Woosley, S.E., \& Weaver, T.A. 1995, ApJS, 101, 181 (WW1995)
\bibitem[1989]{wyse89}Wyse, R.F.G., \& Silk, J. 1989, ApJ, 339, 700

\end{thebibliography}
\end{document}